\documentclass[%
preprint,
 amsmath,amssymb,
 aps,
]{revtex4-1}

\usepackage{graphicx}
\usepackage{dcolumn}
\usepackage{bm}
\usepackage{hyperref}
\usepackage[mathlines]{lineno}
\usepackage{float}

\begin{document}

\preprint{APS/123-QED}

\title{Predicting Scattering Scanning Near-field Optical Microscopy of\\Mass-produced Plasmonic Devices}

\author{Lauren M. Otto}
\affiliation{
  Electrical and Computer Engineering, University of Minnesota, Minneapolis, MN, USA
  }
\affiliation{
  HGST (Western Digital Corporation), San Jose \& Fremont, CA, USA
  }

\author{Stanley P. Burgos}
\affiliation{
  HGST (Western Digital Corporation), San Jose \& Fremont, CA, USA
  }

\author{Matteo Staffaroni}
\affiliation{
  HGST (Western Digital Corporation), San Jose \& Fremont, CA, USA
  }
  
\author{Shen Ren}
\affiliation{
  HGST (Western Digital Corporation), San Jose \& Fremont, CA, USA
  }
  
\author{\"Ozg\"un S\"uzer}
\affiliation{
  HGST (Western Digital Corporation), San Jose \& Fremont, CA, USA
  }
  
\author{Barry C. Stipe}
\affiliation{
  HGST (Western Digital Corporation), San Jose \& Fremont, CA, USA
  }
  
\author{Paul D. Ashby}
\affiliation{
  Molecular Foundry, Lawrence Berkeley National Laboratory, Berkeley, CA, USA
  }

\author{Aeron T. Hammack}%
 \email{aeronth@berkeley.edu}
\affiliation{
  Molecular Foundry, Lawrence Berkeley National Laboratory, Berkeley, CA, USA
  }
\affiliation{
  HGST (Western Digital Corporation), San Jose \& Fremont, CA, USA
  }
  
\date{\today}

\begin{abstract}
  Scattering scanning near-field optical microscopy enables optical imaging and characterization of plasmonic devices with nanometer-scale resolution well below the diffraction limit. This technique enables developers to probe and understand the waveguide-coupled plasmonic antenna in as-fabricated heat-assisted magnetic recording heads. In order validate and predict results and to extract information from experimental measurements that is physically comparable to simulations, a model was developed to translate the simulated electric field into expected near-field measurements using physical parameters specific to scattering scanning near-field optical microscopy physics. The methods used in this paper prove that scattering scanning near-field optical microscopy can be used to determine critical sub-diffraction-limited dimensions of optical field confinement, which is a crucial metrology requirement for the future of nano-optics, semiconductor photonic devices, and biological sensing where the near-field character of light is fundamental to device operation.
\end{abstract}

\maketitle


\section{\label{sec:intro}Introduction}
Today’s optic, photonic, and plasmonic devices incorporate light into nanoscale devices that enable diverse new functionality for chemical or biological sensing and imaging \cite{Jeanmaire1977}, precision light sources \cite{Painter1999,Oulton2009,Zhu2015}, silicon photonics-based communications and signal processing \cite{Yariv2007}, and data storage \cite{Stipe2010,Challener2009}. These devices are often critically dependent on the behavior of light well below the diffraction limit, and therefore require robust near-field characterization techniques, linked with numerical predictions and validations, to design and fabricate functioning devices. This requirement becomes even more essential when such devices are to be manufactured in massive quantities that require process quality control. In fact, future hard disk drives (HDDs) will rely on near-field optical heaters for data storage, and the successful development and production of novel HDD magneto-optical write heads is largely contingent on the ability to reliably predict and validate the near-field behavior of as fabricated devices by direct experimental characterization. The technical improvements to scattering scanning near-field optical microscopy that we have developed for HDD write head optical antenna characterization are potentially extensible and beneficial to any field of study involving the direct optical measurement of photonic modes arising in proximity to nanoscale features.

\section{\label{sec:HAMR_background}Heat Assisted Magnetic Recording Background}
Despite alternative technology developments, most of the world's increasing data storage demands are still satisfied by HDDs. To improve efficiency, the HDD industry has encountered and worked to overcome fundamental limitations preventing the further areal density growth on a disk of magnetic material \cite{Stipe2010,Challener2009,Lu1994,Weller1999,Zhou2014a}. Currently used magnetic media cannot support smaller bit sizes (known as the superparamagnetic limit, smaller bits are less stable), but magnetically harder media has proven too difficult to write (alter magnetization) using traditional magnetic write head strengths/capabilities. In order for data to be recorded into this media, a local spot on the disk with dimensions of the desired bit size (10s of nanometers) must be heated close to its Curie temperature resulting in a significant drop of its magnetic coercivity, thus allowing the magnetic polarization to be switched under the applied field. The chosen method for locally heating the magnetic media in under-development devices is a plasmonic antenna, which focuses far-field optical light into high-intensity and rapidly decaying near-field optical energy, which is directed onto the recording medium (Figure \ref{figure1}a). The high-intensity near-field energy is capable of heating the magnetic media (only a few nanometers away) to the desired temperature while the rapid decay of the evanescent field ensures that the thermal gradient in the media is large so the resulting bit size is comparable to the size of the antenna and its near-field ``spot.'' During drive operation the media is a fast-rotating disk, and as the disk rotates, the magnetic grains in the heated spot experience an applied magnetic field from the write pole, thereby switching their magnetization. Upon further disk rotation, the recorded bit is removed from the near-field hot spot, cools, and the magnetization is therefore ``frozen'' into the media. This technology is known as heat-assisted magnetic recording (HAMR), and it is capable of achieving the desired smaller bit sizes and higher areal densities \cite{Stipe2010,Challener2009}.

Designing, developing, and fabricating the HAMR technology has numerous challenges associated with the newly integrated laser, photonic, and plasmonic elements in addition to the magnetic elements already present. While electromagnetic simulations are immensely powerful and able to guide design and predict results, experimental characterization techniques are necessary to verify modeling results when compared to empirical observations of as-fabricated devices, which can be subjected to many variations, and to determine failure mechanisms. In HAMR, the introduction of a plasmonic antenna generating a deep subwavelength near-field spot calls for a characterization technique capable of investigating the properties and consistency of this crucial feature of the HAMR head’s performance. Former work has demonstrated scattering scanning near-field optical microscopy (sSNOM) to be a top candidate for this task \cite{Zhou2014a,Hillenbrand2000,Dorfmuller2009,Imura2004,Taubner2005,Rang2008,Esteban2008,Schnell2010,Zhou2014}.

\section{\label{sec:sSNOM_background}Scattering Scanning Near-field Optical Microscopy Background}
SNOM was originally developed as a probe system with an aperture in a diaphragm \cite{Ash1972,Lewis1984}, which later was traded for a tapered optical fiber coated with metal such as gold, silver, or aluminum \cite{Pohl1984,Betzig1992,Betzig1993}. Propagating far-field light inside the fiber is converted to near-field radiation by the angled metal coating and then illuminates the sample. This probe also serves to convert the sample-interacted near-field back into propagating far-field light, which would be observed back through the optical fiber. However, the resolution limit of apertured SNOM is determined by the dimensions of the aperture and the excitation wavelength so that as the aperture shrinks, the amount of electromagnetic energy coupled through the aperture and transmitted to the far-field falls off precipitously according to the Bethe limit ($\propto a^6/ \lambda ^{2}$) \cite{Bethe1944}. (Bethe's analytical near-the-hole solutions were later corrected by Bouwkamp \cite{Bouwkamp1950}, and a modern discussion is provided by Novotny \textit{et al.} \cite{Novotny2012}) As a result, for small aperture sizes, the near-field signal accessible using an aperture probe falls below the background and noise thresholds for the technique, thus limiting resolution of apertured SNOM to $\sim$50 nm for visible or near-infrared light. Because the HAMR plasmonic antennas are generally smaller than this size, a technique offering better resolution is necessary.

In sSNOM, the resolution of the SNOM technique is increased by incorporating an atomic force microscopy (AFM) scan probe tip that scatters, rather than transmits, the near-field light into the far-field, thus causing the sharpness of the AFM probe to determine the resolution rather than the size of the aperture and the wavelength \cite{Wickramasinghe1990,Zenhausern1995,Lahrech1996,Knoll1999}. For sub-ten-nanometer tip radii, the scattering cross-section of light is still more than sufficient to beat the signal-to-noise and signal-to-background limits when using a lock-in amplifier to detect the light scattered from the tip \cite{Hillenbrand2000}. In the case of metallic or metal-coated tips, the tip both generates the probing near-field (through the ``lightning rod'' or ``nanofocusing'' effect)\cite{Martin2001,Stockman2004} and performs the function of a dipole scatterer by converting the sample-interacted near-field radiation back into far-field light that can be collected by an objective \cite{Zenhausern1995,Lahrech1996,Knoll1999}. However, for HAMR plasmonic antennas (as well as for other general plasmonic structures), the near-field is generated by the sample (excited by an external source), and so tips with minimal signal disruption while still maintaining high scattering cross-sections are desired \cite{Dorfmuller2009}. In this work, uncoated silicon tips with a typical radius of $\sim$5-15 nm and optically accessible tip apexes were used.

\begin{figure}[H]
\begin{centering}
	\includegraphics[width=1\textwidth]{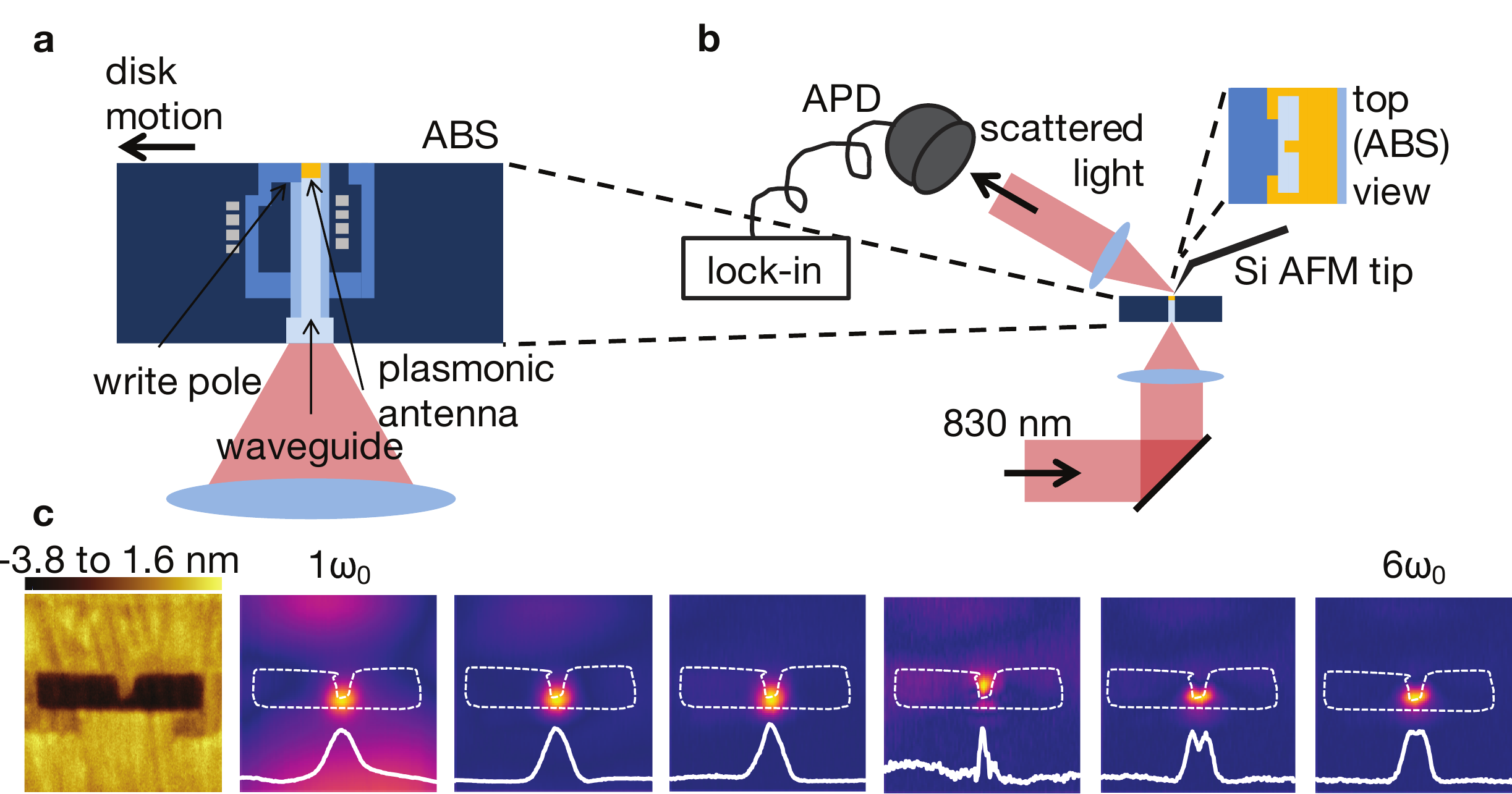}
	\caption[Experimental layout and typical results.]{\textbf{Experimental layout and typical results.} (a) A side profile schematic of a HAMR head including the added elements of a waveguide-coupled laser source which illuminates a plasmonic antenna located on the ABS. During operation, the ABS and plasmonic antenna are positioned only a few nanometers from the media surface. (b) Side profile schematic of the sSNOM system showing the methods of illumination with an 830 nm laser and detection of AFM tip scattered near-field light with an avalanche photodiode (APD) and lock-in amplifier. (c) Plasmonic antenna AFM image together with near-field harmonic mappings collected from the first six harmonics of the AFM tip cantilever resonant frequency. The AFM image as well as the first three harmonics were collected simultaneously while the next three were collected in a second set of scans. Maps are 400 nm $\times$ 400 nm.}
	\label{figure1}
\end{centering}
\end{figure}

Given the large amount of background scattered light that is generally present in sSNOM measurements, the signal specific to the tip apex can be extracted through the use of a lock-in amplifier tuned to the fundamental and higher harmonics of the AFM tip cantilever’s resonant oscillation frequency \cite{Hillenbrand2000,Bek2006}. Prior work has demonstrated that the lower order harmonics include both the propagating and evanescent components of the scattered light, while the higher order harmonics have a stronger dependence on the near-field strength and hence contain higher contrast near-field information \cite{Hillenbrand2000}. The sSNOM system used to measure the HAMR plasmonic antennas (Figure \ref{figure1}b) in this work uses this lock-in driven technique and is capable of capturing up to three harmonics simultaneously during AFM measurements (Figure \ref{figure1}c).

\section{\label{sec:predicting}Predicting Behavior of Mass-produced Plasmonic Devices}

\begin{figure}[H]
\begin{centering}
	\includegraphics[width=1\textwidth]{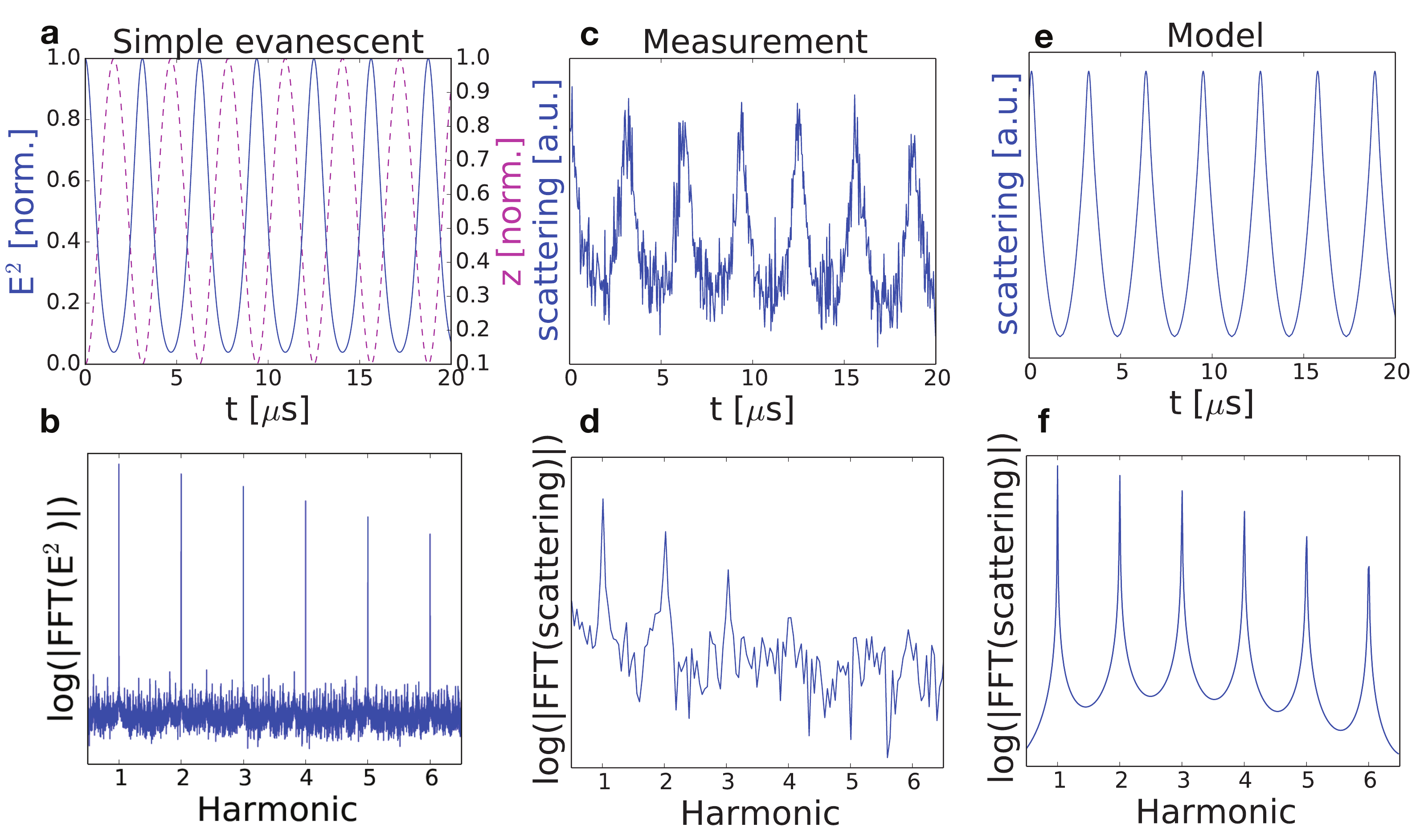}
	\caption[sSNOM model theory]{\textbf{Examining approximations for the measured harmonics.} Comparison of (a,c,e) scattering signal and (b,d,f) harmonics derived from the fast Fourier transform of the corresponding scattering signal for three situations: (a,b) general approximation where the tip height (z) above the ABS varies sinsoidally in time with frequency $\omega _{0}$ and the electric field ($E$) decays exponentially with $z$, (c,d) measurements from the sSNOM system with the tip positioned over the plasmonic antenna, and (e,f) modeled scenarios as calculated using simulated electric field data and accounting for the tip’s effects.}
	\label{sSNOMHAMRmodeltheory}
\end{centering}
\end{figure}

In order to use sSNOM to guide the design and development of HAMR devices, we developed a model to map the electric field data from electromagnetic simulations into the expected harmonic mappings generated by the sSNOM system. The first stage began with a few general approximations. It was assumed that the position of the AFM tip varied sinusoidally in time with the resonant frequency of the cantilever ($z \propto sin(\omega _{0} t)$), that the evanescent electric field followed an exponential decay with increasing distance from sample surface $z$ ($E \propto e^{-z}$), and that the scattering was proportional to intensity ($\sigma \propto E^2$). As expected, the model yields maximum scattering intensity when the tip is closest to the surface (Figure \ref{sSNOMHAMRmodeltheory}a, here $\omega _{0}$ = 300 kHz), and a fast Fourier transform (FFT) of the scattered field yields harmonics that decay exponentially in intensity with increasing order. Using the sSNOM system and oscilloscope capabilities of the lock-in amplifier software, a scattering signal similar in shape to that of the simple evanescent approximation was observed when the tip was hovered (oscillating) over the center feature of the plasmonic antenna containing the most intense near-field signal (Figure \ref{sSNOMHAMRmodeltheory}c). Despite the noise present in this data, a numerical FFT still reveals several harmonics (Figure \ref{sSNOMHAMRmodeltheory}d). It is worth noting that this FFT is performed through numerical analysis of the captured time-domain oscilloscope trace, not by the lock-in amplifier hardware. The lock-in harmonic acquisition during imaging was performed with sufficiently long time constants in order to improve the signal-to-noise ratio in the higher harmonics significantly above the threshold apparent in this post-processed FFT trace.

Because the sSNOM system and HAMR devices are considerably more complex than the simple evanescent model, further improvements to the model were made to better match experimental conditions. To substitute for the simple evanescent electric field assumption ($E \propto e^{-z}$), data from electromagnetic simulations were introduced to account for the actual field expected for the as-designed dimensions of the HAMR plasmonic antenna. The tip-sample interaction was then accounted for in two stages. The first stage weights each 1 nm $\times$ 1 nm $z$-column (where the column is aligned normal to the surface of the sample) of simulated electric field data by a scattering parameter $\sigma$: \cite{Schnell2010,Knoll1999,Knoll2000,Bouhelier2003}
\begin{equation}
	E_{sSNOM} = \sigma E_{model}
	\label{sSNOMfield}
\end{equation}
which yields the scattering parameter adjusted electric field. The scattering parameter $\sigma$ is derived from a spherical probe and image dipole interaction and is defined by the following set of equations: \cite{Ash1972,Zenhausern1995,Knoll1999} 
\begin{equation}
\begin{aligned}
	\label{sSNOMscattering}
	\sigma &= \alpha _{eff} \frac{\pi \sqrt{\frac{8\pi}{3}}}{\lambda ^{2}} \\
	\alpha _{eff} &= \frac{\alpha (1 + \beta)}{1 - \frac{\alpha \beta}{16 \pi (a + z)^{2}}} \\
	\alpha &= 4\pi a^3 \frac{\epsilon _t - 1}{\epsilon _t +2} \\
	\beta &= \frac{\epsilon _s -1}{\epsilon _s +1} \\
\end{aligned}
\end{equation}
where $\alpha _{eff}$ is the effective polarizability of the tip-sample combination, $\alpha$ is the polarizability of the silicon tip (approximated here as a sphere), $\lambda$ is the free space wavelength of the incident light, and $\epsilon _{t}$ and $\epsilon _{s}$ are the complex dielectric functions of the tip and sample, respectively. $\epsilon _{s}$ was allowed to vary based on the material (either gold, oxide, or magnetic material) located directly below the $z$-column being computed.

After applying the scattering transformation to the electric field, the second stage accounted for the shape of the tip. Planar slices of the resulting weighted electric field parallel to the air-bearing surface (ABS), which is the surface facing the rotating magnetic media containing the plasmonic antenna (in the $xy$ plane), were serially convolved with two matrices with each representing different components of the shape of the AFM tip. One matrix was designed as a cone to represent the shape of the tip’s shaft, and the other matrix was designed as a top-hat to represent the shape of the end of the tip. After these convolutions, for a 1 nm $\times$ 1 nm $z$-column positioned directly above the plasmonic antenna, it can be seen that the modeled time-dependent scattered field (Figure \ref{sSNOMHAMRmodeltheory}e) more closely matches the shape of the measured scattered field of the sSNOM signal (Figure \ref{sSNOMHAMRmodeltheory}c) than does the scattered field resulting from the simple evanescent model (Figure \ref{sSNOMHAMRmodeltheory}a). Similarly, a FFT of this modeled sSNOM signal (Figure \ref{sSNOMHAMRmodeltheory}f) mimics the behavior of the measured signal from the lock-in amplifier (Figure \ref{sSNOMHAMRmodeltheory}d) yielding smooth harmonic spikes, and when performed over the entire $xy$ plane, yields the desired modeled near-field harmonic mappings.

\section{\label{sec:tip_effects}Effects of the AFM Tip's Shape}

\begin{figure}
\begin{centering}
	\includegraphics[width=0.8\textwidth]{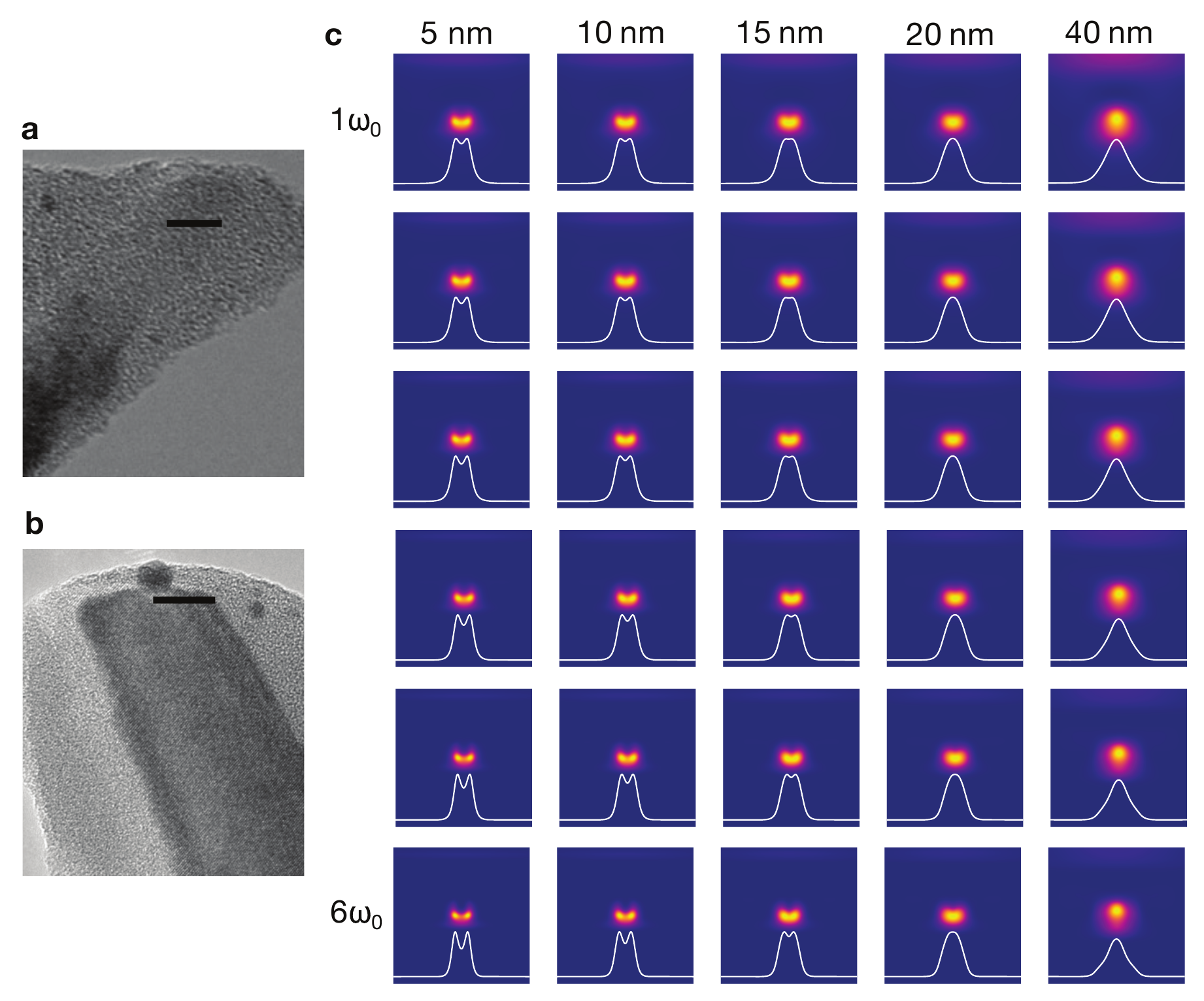}
	\caption[Tip radius effects]{\textbf{Tip radius effects.} A transmission electron microscopy (TEM) image of (a) a better-than-average $\sim$ 5 nm radius tip and (b) an average $\sim$ 10 nm tip, both coated with material (likely from the head) after scanning. Scale bars 5 nm (a) and 10 nm (b), respectively. (c) Near-field maps of the modeled system as a function of tip radius and harmonic assuming the silicon portion of the tip has a distance of closest approach (DCA) to the sample of 8 nm. Near-field maps are 400 nm $\times$ 400 nm.}
	\label{figure3}
\end{centering}
\end{figure}

Development of the sSNOM model required further investigation of the AFM tip’s local shape interacting with the high-intensity near-field. Based on manufacturer specifications, our experience, and previous transmission electron microscopy (TEM) images, it was assumed that most tips used in these experiments contained a 5-15 nm radius, where the sharper tips (5 nm radius, Figure \ref{figure3}a) gave a sSNOM signal with better resolution than average (10 nm radius, Figure \ref{figure3}b) or blunter ($>$15 nm radius) tips. Furthermore, TEMs of tips imaged post-scanning revealed several nanometers of material built up on the tip, including in the region separating the silicon tip from the sample. Therefore, the corresponding change in tip-sample separation was accounted for in the sSNOM model in addition to the tip’s radius. The TEM images also revealed the angled nature of the used tip’s shaft as well as the flat top-hat nature of its end. 

Both the shape of the tip’s shaft as well as its end were incorporated into the sSNOM model through the two convolution matrices described above. The shaft was approximated as a cone with a half-angle of 9$^\circ$, and the tip end was approximated as a top-hat (step function in cylindrical coordinates) with variable radius. Radii of 5, 10, and 15 nm represented the range of usual tips while radii of 20 and 40 nm represented more extreme cases of blunt tips (Figure \ref{figure3}c). Based on the developed sSNOM model and previous experience with these tips, a radius of 15 nm was used in the model for correlation with the experimental data. Further improvement of the model is possible through the use of a three-dimensional volumetric convolution, rather than two-dimensional planar convolution used in this work, of the conical tip shaft matrix (including the tip’s angled rather than vertical approach) with the scattered field, which is expected to more accurately model the physical configuration and may increase the far-field contribution to the $1 \omega _{0}$. 


\section{\label{sec:closest_approach_effects}Effects of the AFM Tip's Distance of Closest Approach}

\begin{figure}
\begin{centering}
	\includegraphics[width=0.7\textwidth]{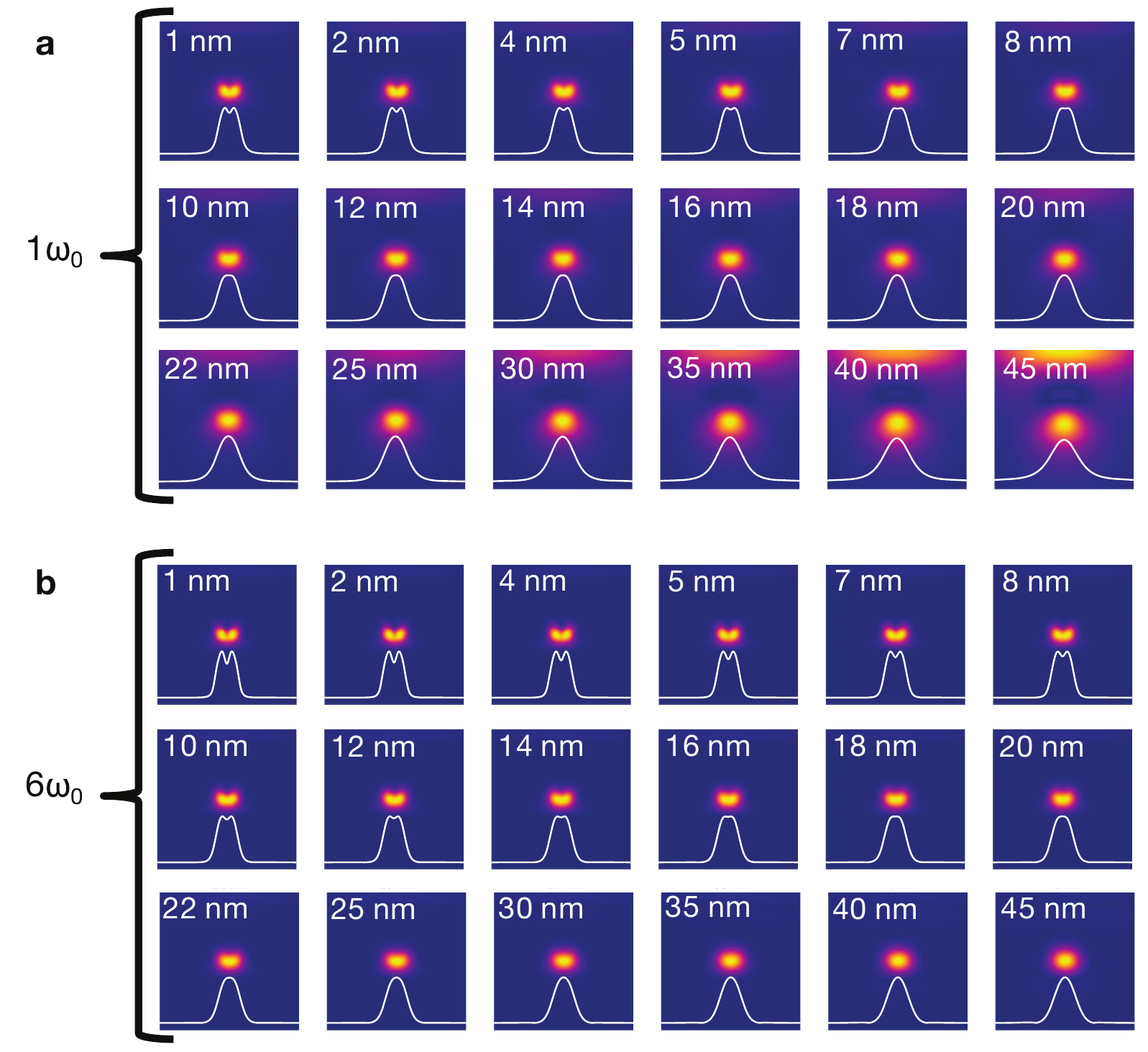}
	\caption[Maximum Intensity vs. Tip Radius]{\textbf{Modeled near-field data as a function of DCA and harmonic assuming a 15 nm tip radius.} (a) At the tip cantilever’s resonant frequency, $1 \omega _{0}$. (b) At $6 \omega _{0}$. Near-field maps are 400 nm $\times$ 400 nm.}
	\label{figure4}
\end{centering}
\end{figure}

Further investigation of the mechanical AFM behavior of the sSNOM system was performed in order to determine the tip’s distance of closest approach (DCA) to the sample. In the sSNOM model, the DCA value was set as the minimum $z$ height (and cutoff) for the FFTs of the scattered signal. It was found that the chosen DCA had a significant effect on the sSNOM model’s results (Figure \ref{figure4}). A smaller DCA (tip approaches much closer to the ABS) incorporates much more of the dominant near-field signal (as opposed to the weaker far-field signal) into the FFT, which becomes evident in the harmonic maps for the fundamental frequency ($1 \omega _{0}$, Figure \ref{figure4}a) when compared with the higher $6 \omega _{0}$ frequency (Figure \ref{figure4}b). Maximum intensity plots are also included for all investigated harmonics and heights as well as additional harmonic maps in the Supplemental Information (Figures \ref{SIHarm815}-\ref{SIMItiprad}).

Estimation of tip DCA was also performed with a mechanical model of tapping mode AFM, which solves the equation of motion for the cantilever numerically as it encounters tip-sample forces modeled to include van der Waals attraction, capillary adhesion due to a water layer, and Pauli hard sphere repulsion \cite{Chen2004}. When inputting the tip radius, free amplitude, and amplitude setpoint (see Supplementary Information), and using the measured phase signal as a check for accuracy, the model shows that the cantilever was oscillating in the repulsive regime indicating that the tip approaches the surface very closely and experiences repulsive contact with the sample at the bottom of most oscillations.

\section{\label{sec:results}Results}

Accounting for the thickness of material collected by the tips during scans ($\sim$5 nm in Figure \ref{figure3}a,b) as well as other complex effects due to the water meniscus between the tip and the sample surface \cite{Chen2004,Calleja2002}, and comparing the experimentally measured data with the sSNOM model’s results, a DCA of 8 nm was chosen together with a tip radius of 15 nm to represent the best fit of the sSNOM model to the experimental measurements and expectations. A ``good fit'' between the harmonic maps generated by the sSNOM model and the experimental data was established when a bifurcated near-field signal was observed in the higher order harmonics ($5 \omega _{0}$, $6 \omega _{0}$) but not in the lower order harmonics ($1-3 \omega _{0}$). The visible near-field bifurcation requirement for the higher harmonics ruled out the possibility of the experimental tip having a radius of 20 nm or larger since no bifurcation was evident, even in the $6 \omega _{0}$ (Figure \ref{SIHarm20}), when the chosen distance of closest approach was 1 nm (the closest possible in the present model). In addition, good qualitative agreement between the experimental and modeled near-field maps using smaller tip radii ($<$15 nm) required DCA values exceeding expectations and so were ruled out by the AFM mechanical model (Figures \ref{SIHarm10}-\ref{SIHarm12nmDCA}). Since the sSNOM model’s results show significant dependence on the DCA, incorporating simulated electric field data with topographic sensitivity (the presented simulations assumed a planar surface) would more closely match the experimental conditions revealed by the recessed nature of the waveguide relative to the antenna seen in the AFM image and could improve the model.

\begin{figure}[H]
\begin{centering}
	\includegraphics[width=1\textwidth]{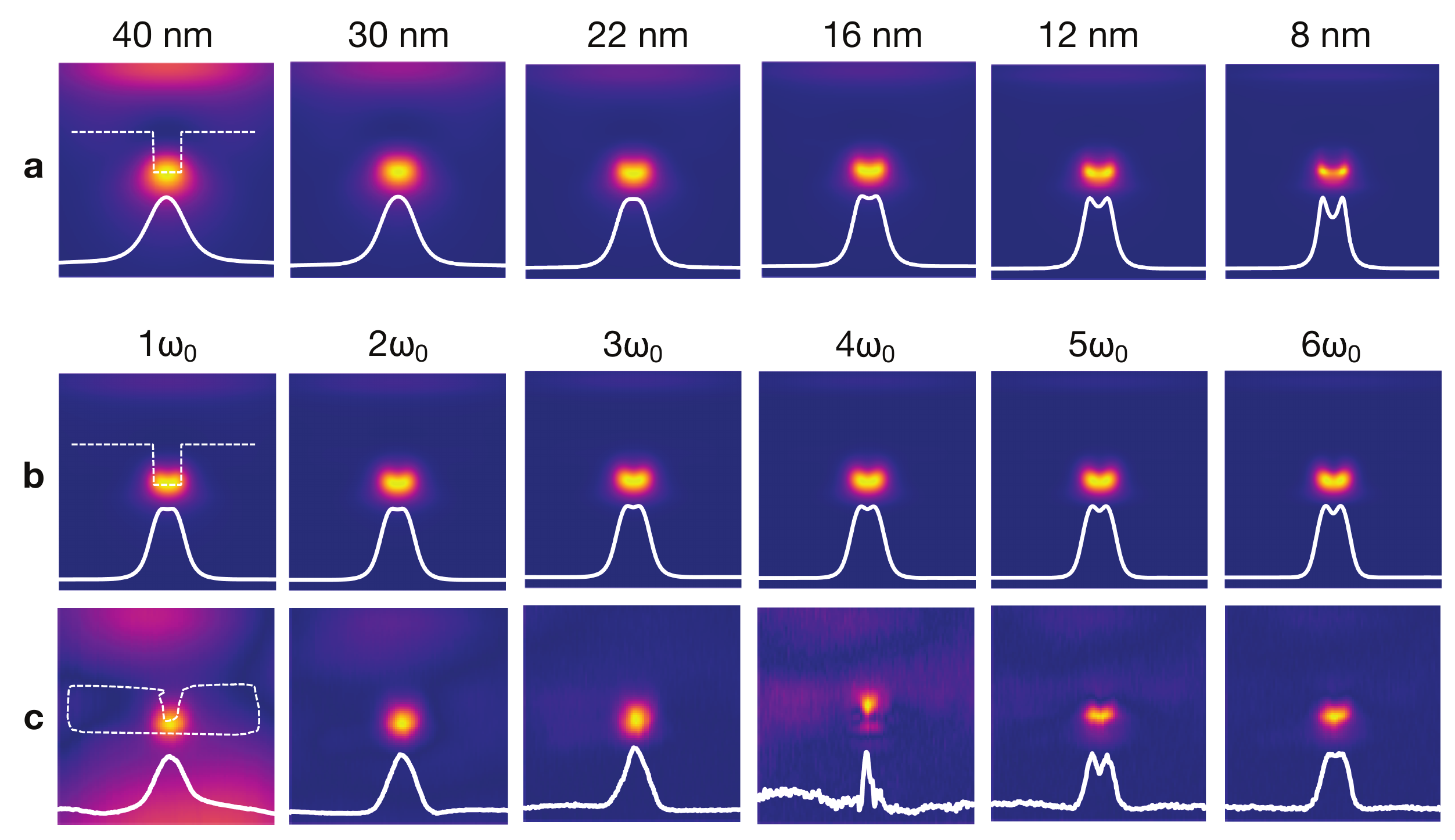}
	\caption[Result of sSNOM model compared to measurements]{\textbf{Final result of model derived from simulated data and compared to measured data.} Comparison of a. simulated electric field intensity, b. modeled near-field images, and c. sSNOM measured near-field images. All maps are 400 nm $\times$ 400 nm.}
	\label{sSNOMHAMRfinalresult}
\end{centering}
\end{figure}

Slices of the simulated electric field intensity at different heights of interest (Figure \ref{sSNOMHAMRfinalresult}a, with further details included in the Supplemental Information are compared to the corresponding harmonics from the sSNOM model results using a tip radius of 15 nm and a DCA of 8 nm (Figure \ref{sSNOMHAMRfinalresult}b) and the experimentally measured harmonic maps (Figure \ref{sSNOMHAMRfinalresult}c). From this comparison, it can be seen that higher harmonics, such as $5 \omega _{0}$ and $6 \omega _{0}$, are more representative of the near-field in the region that would be occupied by the magnetic media during drive operation than their lower order harmonic counterparts. The nature of the higher order harmonics in which this relationship was found signified the prescribed rapid change in near-field intensity closer to the plasmonic antenna. The strange behavior exhibited by the measured $4 \omega _{0}$ was likely caused by a non-sinusoidal feature at the bottom of the tip’s oscillation due the tip’s possible contact with and/or repulsion by the sample surface during its motion. The non-sinusoidal motion of the tip was not accounted for in the present sSNOM model, but it could be introduced upon further mechanical characterization of the sSNOM-tip system and its interaction with the sample. An introduction of a sharp change in the peak scattering would surely become evident in one or more of the higher order harmonics in the presence of non-sinusoidal tip oscillation. In the case of this experiment, this feature was found in the $4 \omega _{0}$, and it is also expected to have been present in the $8 \omega _{0}$, etc. This observation as well as the AFM mechanical model results and the flat nature of the post-scanning tips suggests that the scan parameters cause significant tip interaction with the sample likely leading to the tip’s top-hat shape. Scan parameters for typical non-contact tip operation involve small oscillation amplitudes that fail to sweep sufficient evanescent field variation and are not capable of performing the desired full characterization of the plasmonic antenna’s near-field. For this reason, large oscillation amplitudes were used that resulted in greater tip-sample interactions. 

In conclusion, sSNOM is a useful technique capable of characterizing HAMR heads for next-generation hard disk drives. The development of a sSNOM model to translate from simulated electric field data was necessary in order to derive meaning from the measured near-field harmonics. Furthermore, the same model approach can be applied to other plasmonic structures across many disciplines and applications.

\section{\label{sec:sSNOMMethods}Methods}
Steady state electromagnetic modeling of the plasmonic system was performed using the FDTD method in Lumerical.

sSNOM measurements were performed with the AIST-NT CombiScope 1000-SPM and tips from Nanosensors (ATEC-NC). The auto non-contact/tapping default mode was used with the 160 $\mu$m cantilevers ($\sim$300 kHz resonance), 80 nm oscillation amplitude, and 87 percent set point. An 830 nm laser (diode: Sanyo DL8142-201) was polarized (Thorlabs LPVIS100-MP) and directed with free space optics into the microscope. The microscope’s bottom objective (Olympus ULWD MSPlan 50) mount was piezo tuned in three dimensions first for coarse alignment and then for optimal coupling of the laser spot into the waveguide once the tip was hovered over the plasmonic antenna. A 50$\times$ objective (Mitutoyo NIR M Plan Apo NIR 50$\times$, 378-825-5) was precisely aligned with piezo scanners (in objective’s $xy$ plane, manual in $z$) and used to collect the scattered light, which was then observed by an APD (Thorlabs APD120A). The resulting signal passed through a lock-in amplifier (Zurich Instruments HF2LI), and data was collected through the AIST-NT software. 

\section{\label{sec:Acknowledgements}Acknowledgements}
L.M.O. acknowledges the National Institutes of Health Biotechnology Training Grant (T 32 GM008347) and the National Science Foundation Graduate Research Fellowship Program (00039202). Work performed at the Molecular Foundry (TEM images and AFM mechanical model) was supported by the Office of Science, Office of Basic Energy Sciences, of the U.S. Department of Energy under contract no. DE-AC02-05CH11231.

\bibliography{sSNOM}

\begin{thebibliography}{37}%
\makeatletter
\providecommand \@ifxundefined [1]{%
 \@ifx{#1\undefined}
}%
\providecommand \@ifnum [1]{%
 \ifnum #1\expandafter \@firstoftwo
 \else \expandafter \@secondoftwo
 \fi
}%
\providecommand \@ifx [1]{%
 \ifx #1\expandafter \@firstoftwo
 \else \expandafter \@secondoftwo
 \fi
}%
\providecommand \natexlab [1]{#1}%
\providecommand \enquote  [1]{``#1''}%
\providecommand \bibnamefont  [1]{#1}%
\providecommand \bibfnamefont [1]{#1}%
\providecommand \citenamefont [1]{#1}%
\providecommand \href@noop [0]{\@secondoftwo}%
\providecommand \href [0]{\begingroup \@sanitize@url \@href}%
\providecommand \@href[1]{\@@startlink{#1}\@@href}%
\providecommand \@@href[1]{\endgroup#1\@@endlink}%
\providecommand \@sanitize@url [0]{\catcode `\\12\catcode `\$12\catcode
  `\&12\catcode `\#12\catcode `\^12\catcode `\_12\catcode `\%12\relax}%
\providecommand \@@startlink[1]{}%
\providecommand \@@endlink[0]{}%
\providecommand \url  [0]{\begingroup\@sanitize@url \@url }%
\providecommand \@url [1]{\endgroup\@href {#1}{\urlprefix }}%
\providecommand \urlprefix  [0]{URL }%
\providecommand \Eprint [0]{\href }%
\providecommand \doibase [0]{http://dx.doi.org/}%
\providecommand \selectlanguage [0]{\@gobble}%
\providecommand \bibinfo  [0]{\@secondoftwo}%
\providecommand \bibfield  [0]{\@secondoftwo}%
\providecommand \translation [1]{[#1]}%
\providecommand \BibitemOpen [0]{}%
\providecommand \bibitemStop [0]{}%
\providecommand \bibitemNoStop [0]{.\EOS\space}%
\providecommand \EOS [0]{\spacefactor3000\relax}%
\providecommand \BibitemShut  [1]{\csname bibitem#1\endcsname}%
\let\auto@bib@innerbib\@empty
\bibitem [{\citenamefont {Jeanmaire}\ and\ \citenamefont {{Van
  Duyne}}(1977)}]{Jeanmaire1977}%
  \BibitemOpen
  \bibfield  {author} {\bibinfo {author} {\bibfnamefont {D.~L.}\ \bibnamefont
  {Jeanmaire}}\ and\ \bibinfo {author} {\bibfnamefont {R.~P.}\ \bibnamefont
  {{Van Duyne}}},\ }\href@noop {} {\bibfield  {journal} {\bibinfo  {journal}
  {Journal of Electroanalytical Chemistry}\ }\textbf {\bibinfo {volume} {84}},\
  \bibinfo {pages} {1} (\bibinfo {year} {1977})}\BibitemShut {NoStop}%
\bibitem [{\citenamefont {Painter}\ \emph {et~al.}(1999)\citenamefont
  {Painter}, \citenamefont {Lee}, \citenamefont {Scherer}, \citenamefont
  {Yariv}, \citenamefont {O'Brien}, \citenamefont {Dapkus},\ and\ \citenamefont
  {Kim}}]{Painter1999}%
  \BibitemOpen
  \bibfield  {author} {\bibinfo {author} {\bibfnamefont {O.}~\bibnamefont
  {Painter}}, \bibinfo {author} {\bibfnamefont {R.~K.}\ \bibnamefont {Lee}},
  \bibinfo {author} {\bibfnamefont {A.}~\bibnamefont {Scherer}}, \bibinfo
  {author} {\bibfnamefont {A.}~\bibnamefont {Yariv}}, \bibinfo {author}
  {\bibfnamefont {J.~D.}\ \bibnamefont {O'Brien}}, \bibinfo {author}
  {\bibfnamefont {P.~D.}\ \bibnamefont {Dapkus}}, \ and\ \bibinfo {author}
  {\bibfnamefont {I.}~\bibnamefont {Kim}},\ }\href@noop {} {\bibfield
  {journal} {\bibinfo  {journal} {Science}\ }\textbf {\bibinfo {volume}
  {284}},\ \bibinfo {pages} {1819} (\bibinfo {year} {1999})}\BibitemShut
  {NoStop}%
\bibitem [{\citenamefont {Oulton}\ \emph {et~al.}(2009)\citenamefont {Oulton},
  \citenamefont {Sorger}, \citenamefont {Zentgraf}, \citenamefont {Ma},
  \citenamefont {Gladden}, \citenamefont {Dai}, \citenamefont {Bartal},\ and\
  \citenamefont {Zhang}}]{Oulton2009}%
  \BibitemOpen
  \bibfield  {author} {\bibinfo {author} {\bibfnamefont {R.~F.}\ \bibnamefont
  {Oulton}}, \bibinfo {author} {\bibfnamefont {V.~J.}\ \bibnamefont {Sorger}},
  \bibinfo {author} {\bibfnamefont {T.}~\bibnamefont {Zentgraf}}, \bibinfo
  {author} {\bibfnamefont {R.-M.}\ \bibnamefont {Ma}}, \bibinfo {author}
  {\bibfnamefont {C.}~\bibnamefont {Gladden}}, \bibinfo {author} {\bibfnamefont
  {L.}~\bibnamefont {Dai}}, \bibinfo {author} {\bibfnamefont {G.}~\bibnamefont
  {Bartal}}, \ and\ \bibinfo {author} {\bibfnamefont {X.}~\bibnamefont
  {Zhang}},\ }\href {\doibase 10.1038/nature08364} {\bibfield  {journal}
  {\bibinfo  {journal} {Nature}\ }\textbf {\bibinfo {volume} {461}},\ \bibinfo
  {pages} {629} (\bibinfo {year} {2009})},\ \Eprint
  {http://arxiv.org/abs/0509168v1} {arXiv:0509168v1 [quant-ph]} \BibitemShut
  {NoStop}%
\bibitem [{\citenamefont {Zhu}\ \emph {et~al.}(2015)\citenamefont {Zhu},
  \citenamefont {Fu}, \citenamefont {Meng}, \citenamefont {Wu}, \citenamefont
  {Gong}, \citenamefont {Ding}, \citenamefont {Gustafsson}, \citenamefont
  {Trinh}, \citenamefont {Jin},\ and\ \citenamefont {Zhu}}]{Zhu2015}%
  \BibitemOpen
  \bibfield  {author} {\bibinfo {author} {\bibfnamefont {H.}~\bibnamefont
  {Zhu}}, \bibinfo {author} {\bibfnamefont {Y.}~\bibnamefont {Fu}}, \bibinfo
  {author} {\bibfnamefont {F.}~\bibnamefont {Meng}}, \bibinfo {author}
  {\bibfnamefont {X.}~\bibnamefont {Wu}}, \bibinfo {author} {\bibfnamefont
  {Z.}~\bibnamefont {Gong}}, \bibinfo {author} {\bibfnamefont {Q.}~\bibnamefont
  {Ding}}, \bibinfo {author} {\bibfnamefont {M.~V.}\ \bibnamefont
  {Gustafsson}}, \bibinfo {author} {\bibfnamefont {M.~T.}\ \bibnamefont
  {Trinh}}, \bibinfo {author} {\bibfnamefont {S.}~\bibnamefont {Jin}}, \ and\
  \bibinfo {author} {\bibfnamefont {X.-Y.}\ \bibnamefont {Zhu}},\ }\href
  {\doibase 10.1038/NMAT4271} {\bibfield  {journal} {\bibinfo  {journal}
  {Nature Materials}\ }\textbf {\bibinfo {volume} {14}},\ \bibinfo {pages}
  {636} (\bibinfo {year} {2015})}\BibitemShut {NoStop}%
\bibitem [{\citenamefont {Yariv}\ and\ \citenamefont {Yeh}(2007)}]{Yariv2007}%
  \BibitemOpen
  \bibfield  {author} {\bibinfo {author} {\bibfnamefont {A.}~\bibnamefont
  {Yariv}}\ and\ \bibinfo {author} {\bibfnamefont {P.}~\bibnamefont {Yeh}},\
  }\href@noop {} {\emph {\bibinfo {title} {{Photonics}}}},\ \bibinfo {edition}
  {6th}\ ed.\ (\bibinfo  {publisher} {Oxford University Press, Inc.},\ \bibinfo
  {address} {New York, New York},\ \bibinfo {year} {2007})\BibitemShut
  {NoStop}%
\bibitem [{\citenamefont {Stipe}\ \emph {et~al.}(2010)\citenamefont {Stipe},
  \citenamefont {Strand}, \citenamefont {Poon}, \citenamefont {Balamane},
  \citenamefont {Boone}, \citenamefont {Katine}, \citenamefont {Li},
  \citenamefont {Rawat}, \citenamefont {Nemoto}, \citenamefont {Hirotsune},
  \citenamefont {Hellwig}, \citenamefont {Ruiz}, \citenamefont {Dobisz},
  \citenamefont {Kercher}, \citenamefont {Robertson}, \citenamefont
  {Albrecht},\ and\ \citenamefont {Terris}}]{Stipe2010}%
  \BibitemOpen
  \bibfield  {author} {\bibinfo {author} {\bibfnamefont {B.~C.}\ \bibnamefont
  {Stipe}}, \bibinfo {author} {\bibfnamefont {T.~C.}\ \bibnamefont {Strand}},
  \bibinfo {author} {\bibfnamefont {C.~C.}\ \bibnamefont {Poon}}, \bibinfo
  {author} {\bibfnamefont {H.}~\bibnamefont {Balamane}}, \bibinfo {author}
  {\bibfnamefont {T.~D.}\ \bibnamefont {Boone}}, \bibinfo {author}
  {\bibfnamefont {J.~A.}\ \bibnamefont {Katine}}, \bibinfo {author}
  {\bibfnamefont {J.-L.}\ \bibnamefont {Li}}, \bibinfo {author} {\bibfnamefont
  {V.}~\bibnamefont {Rawat}}, \bibinfo {author} {\bibfnamefont
  {H.}~\bibnamefont {Nemoto}}, \bibinfo {author} {\bibfnamefont
  {A.}~\bibnamefont {Hirotsune}}, \bibinfo {author} {\bibfnamefont
  {O.}~\bibnamefont {Hellwig}}, \bibinfo {author} {\bibfnamefont
  {R.}~\bibnamefont {Ruiz}}, \bibinfo {author} {\bibfnamefont {E.}~\bibnamefont
  {Dobisz}}, \bibinfo {author} {\bibfnamefont {D.~S.}\ \bibnamefont {Kercher}},
  \bibinfo {author} {\bibfnamefont {N.}~\bibnamefont {Robertson}}, \bibinfo
  {author} {\bibfnamefont {T.~R.}\ \bibnamefont {Albrecht}}, \ and\ \bibinfo
  {author} {\bibfnamefont {B.~D.}\ \bibnamefont {Terris}},\ }\href {\doibase
  10.1038/nphoton.2010.90} {\bibfield  {journal} {\bibinfo  {journal} {Nature
  Photonics}\ }\textbf {\bibinfo {volume} {4}},\ \bibinfo {pages} {484}
  (\bibinfo {year} {2010})}\BibitemShut {NoStop}%
\bibitem [{\citenamefont {Challener}\ \emph {et~al.}(2009)\citenamefont
  {Challener}, \citenamefont {Peng}, \citenamefont {Itagi}, \citenamefont
  {Karns}, \citenamefont {Peng}, \citenamefont {Peng}, \citenamefont {Yang},
  \citenamefont {Zhu}, \citenamefont {Gokemeijer}, \citenamefont {Hsia},
  \citenamefont {Ju}, \citenamefont {Rottmayer}, \citenamefont {Seigler},\ and\
  \citenamefont {Gage}}]{Challener2009}%
  \BibitemOpen
  \bibfield  {author} {\bibinfo {author} {\bibfnamefont {W.~A.}\ \bibnamefont
  {Challener}}, \bibinfo {author} {\bibfnamefont {C.}~\bibnamefont {Peng}},
  \bibinfo {author} {\bibfnamefont {A.~V.}\ \bibnamefont {Itagi}}, \bibinfo
  {author} {\bibfnamefont {D.}~\bibnamefont {Karns}}, \bibinfo {author}
  {\bibfnamefont {W.}~\bibnamefont {Peng}}, \bibinfo {author} {\bibfnamefont
  {Y.}~\bibnamefont {Peng}}, \bibinfo {author} {\bibfnamefont {X.}~\bibnamefont
  {Yang}}, \bibinfo {author} {\bibfnamefont {X.}~\bibnamefont {Zhu}}, \bibinfo
  {author} {\bibfnamefont {N.~J.}\ \bibnamefont {Gokemeijer}}, \bibinfo
  {author} {\bibfnamefont {Y.-T.}\ \bibnamefont {Hsia}}, \bibinfo {author}
  {\bibfnamefont {G.}~\bibnamefont {Ju}}, \bibinfo {author} {\bibfnamefont
  {R.~E.}\ \bibnamefont {Rottmayer}}, \bibinfo {author} {\bibfnamefont {M.~A.}\
  \bibnamefont {Seigler}}, \ and\ \bibinfo {author} {\bibfnamefont {E.~C.}\
  \bibnamefont {Gage}},\ }\href {\doibase 10.1038/nphoton.2009.71} {\bibfield
  {journal} {\bibinfo  {journal} {Nature Photonics}\ }\textbf {\bibinfo
  {volume} {3}},\ \bibinfo {pages} {303} (\bibinfo {year} {2009})}\BibitemShut
  {NoStop}%
\bibitem [{\citenamefont {Lu}\ and\ \citenamefont {Charap}(1994)}]{Lu1994}%
  \BibitemOpen
  \bibfield  {author} {\bibinfo {author} {\bibfnamefont {P.-L.}\ \bibnamefont
  {Lu}}\ and\ \bibinfo {author} {\bibfnamefont {S.~H.}\ \bibnamefont
  {Charap}},\ }\href {\doibase 10.1063/1.355609} {\bibfield  {journal}
  {\bibinfo  {journal} {Journal of Applied Physics}\ }\textbf {\bibinfo
  {volume} {75}},\ \bibinfo {pages} {5768} (\bibinfo {year}
  {1994})}\BibitemShut {NoStop}%
\bibitem [{\citenamefont {Weller}\ and\ \citenamefont
  {Moser}(1999)}]{Weller1999}%
  \BibitemOpen
  \bibfield  {author} {\bibinfo {author} {\bibfnamefont {D.}~\bibnamefont
  {Weller}}\ and\ \bibinfo {author} {\bibfnamefont {A.}~\bibnamefont {Moser}},\
  }\href@noop {} {\bibfield  {journal} {\bibinfo  {journal} {IEEE Transactions
  on Magnetics}\ }\textbf {\bibinfo {volume} {35}},\ \bibinfo {pages} {4423}
  (\bibinfo {year} {1999})}\BibitemShut {NoStop}%
\bibitem [{\citenamefont {Zhou}\ \emph
  {et~al.}(2014{\natexlab{a}})\citenamefont {Zhou}, \citenamefont {Xu},
  \citenamefont {Hammack}, \citenamefont {Stipe}, \citenamefont {Gao},
  \citenamefont {Scholz},\ and\ \citenamefont {Gage}}]{Zhou2014a}%
  \BibitemOpen
  \bibfield  {author} {\bibinfo {author} {\bibfnamefont {N.}~\bibnamefont
  {Zhou}}, \bibinfo {author} {\bibfnamefont {X.}~\bibnamefont {Xu}}, \bibinfo
  {author} {\bibfnamefont {A.~T.}\ \bibnamefont {Hammack}}, \bibinfo {author}
  {\bibfnamefont {B.~C.}\ \bibnamefont {Stipe}}, \bibinfo {author}
  {\bibfnamefont {K.}~\bibnamefont {Gao}}, \bibinfo {author} {\bibfnamefont
  {W.}~\bibnamefont {Scholz}}, \ and\ \bibinfo {author} {\bibfnamefont {E.~C.}\
  \bibnamefont {Gage}},\ }\href {\doibase 10.1515/nanoph-2014-0001} {\bibfield
  {journal} {\bibinfo  {journal} {Nanophotonics}\ }\textbf {\bibinfo {volume}
  {3}},\ \bibinfo {pages} {141} (\bibinfo {year}
  {2014}{\natexlab{a}})}\BibitemShut {NoStop}%
\bibitem [{\citenamefont {Hillenbrand}\ and\ \citenamefont
  {Keilmann}(2000)}]{Hillenbrand2000}%
  \BibitemOpen
  \bibfield  {author} {\bibinfo {author} {\bibfnamefont {R.}~\bibnamefont
  {Hillenbrand}}\ and\ \bibinfo {author} {\bibfnamefont {F.}~\bibnamefont
  {Keilmann}},\ }\href {\doibase 10.1103/PhysRevLett.85.3029} {\bibfield
  {journal} {\bibinfo  {journal} {Physical Review Letters}\ }\textbf {\bibinfo
  {volume} {85}},\ \bibinfo {pages} {3029} (\bibinfo {year}
  {2000})}\BibitemShut {NoStop}%
\bibitem [{\citenamefont {Dorfmüller}\ \emph {et~al.}(2009)\citenamefont
  {Dorfmüller}, \citenamefont {Vogelgesang}, \citenamefont {Weitz},
  \citenamefont {Rockstuhl}, \citenamefont {Etrich}, \citenamefont {Pertsch},
  \citenamefont {Lederer},\ and\ \citenamefont {Kern}}]{Dorfmuller2009}%
  \BibitemOpen
  \bibfield  {author} {\bibinfo {author} {\bibfnamefont {J.}~\bibnamefont
  {Dorfmüller}}, \bibinfo {author} {\bibfnamefont {R.}~\bibnamefont
  {Vogelgesang}}, \bibinfo {author} {\bibfnamefont {R.~T.}\ \bibnamefont
  {Weitz}}, \bibinfo {author} {\bibfnamefont {C.}~\bibnamefont {Rockstuhl}},
  \bibinfo {author} {\bibfnamefont {C.}~\bibnamefont {Etrich}}, \bibinfo
  {author} {\bibfnamefont {T.}~\bibnamefont {Pertsch}}, \bibinfo {author}
  {\bibfnamefont {F.}~\bibnamefont {Lederer}}, \ and\ \bibinfo {author}
  {\bibfnamefont {K.}~\bibnamefont {Kern}},\ }\href {\doibase
  10.1021/nl900900r} {\bibfield  {journal} {\bibinfo  {journal} {Nano Letters}\
  }\textbf {\bibinfo {volume} {9}},\ \bibinfo {pages} {2372} (\bibinfo {year}
  {2009})}\BibitemShut {NoStop}%
\bibitem [{\citenamefont {Imura}\ \emph {et~al.}(2004)\citenamefont {Imura},
  \citenamefont {Nagahara},\ and\ \citenamefont {Okamoto}}]{Imura2004}%
  \BibitemOpen
  \bibfield  {author} {\bibinfo {author} {\bibfnamefont {K.}~\bibnamefont
  {Imura}}, \bibinfo {author} {\bibfnamefont {T.}~\bibnamefont {Nagahara}}, \
  and\ \bibinfo {author} {\bibfnamefont {H.}~\bibnamefont {Okamoto}},\ }\href
  {\doibase 10.1021/jp047950h} {\bibfield  {journal} {\bibinfo  {journal}
  {Journal of Physical Chemistry B}\ }\textbf {\bibinfo {volume} {108}},\
  \bibinfo {pages} {16344} (\bibinfo {year} {2004})}\BibitemShut {NoStop}%
\bibitem [{\citenamefont {Taubner}\ \emph {et~al.}(2005)\citenamefont
  {Taubner}, \citenamefont {Keilmann},\ and\ \citenamefont
  {Hillenbrand}}]{Taubner2005}%
  \BibitemOpen
  \bibfield  {author} {\bibinfo {author} {\bibfnamefont {T.}~\bibnamefont
  {Taubner}}, \bibinfo {author} {\bibfnamefont {F.}~\bibnamefont {Keilmann}}, \
  and\ \bibinfo {author} {\bibfnamefont {R.}~\bibnamefont {Hillenbrand}},\
  }\href {\doibase 10.1364/OPEX.13.008893} {\bibfield  {journal} {\bibinfo
  {journal} {Optics Express}\ }\textbf {\bibinfo {volume} {13}},\ \bibinfo
  {pages} {8893} (\bibinfo {year} {2005})}\BibitemShut {NoStop}%
\bibitem [{\citenamefont {Rang}\ \emph {et~al.}(2008)\citenamefont {Rang},
  \citenamefont {Jones}, \citenamefont {Fei}, \citenamefont {Li}, \citenamefont
  {Wiley}, \citenamefont {Younan},\ and\ \citenamefont {Raschke}}]{Rang2008}%
  \BibitemOpen
  \bibfield  {author} {\bibinfo {author} {\bibfnamefont {M.}~\bibnamefont
  {Rang}}, \bibinfo {author} {\bibfnamefont {A.~C.}\ \bibnamefont {Jones}},
  \bibinfo {author} {\bibfnamefont {Z.}~\bibnamefont {Fei}}, \bibinfo {author}
  {\bibfnamefont {Z.~Y.}\ \bibnamefont {Li}}, \bibinfo {author} {\bibfnamefont
  {B.~J.}\ \bibnamefont {Wiley}}, \bibinfo {author} {\bibfnamefont
  {X.}~\bibnamefont {Younan}}, \ and\ \bibinfo {author} {\bibfnamefont {M.~B.}\
  \bibnamefont {Raschke}},\ }\href {\doibase 10.1021/nl801808b} {\bibfield
  {journal} {\bibinfo  {journal} {Nano Letters}\ }\textbf {\bibinfo {volume}
  {8}},\ \bibinfo {pages} {3357} (\bibinfo {year} {2008})}\BibitemShut
  {NoStop}%
\bibitem [{\citenamefont {Esteban}\ \emph {et~al.}(2008)\citenamefont
  {Esteban}, \citenamefont {Vogelgesang}, \citenamefont {Dorfm{\"{u}}ller},
  \citenamefont {Dmitriev}, \citenamefont {Rockstuhl}, \citenamefont {Etrich},\
  and\ \citenamefont {Kern}}]{Esteban2008}%
  \BibitemOpen
  \bibfield  {author} {\bibinfo {author} {\bibfnamefont {R.}~\bibnamefont
  {Esteban}}, \bibinfo {author} {\bibfnamefont {R.}~\bibnamefont
  {Vogelgesang}}, \bibinfo {author} {\bibfnamefont {J.}~\bibnamefont
  {Dorfm{\"{u}}ller}}, \bibinfo {author} {\bibfnamefont {a.}~\bibnamefont
  {Dmitriev}}, \bibinfo {author} {\bibfnamefont {C.}~\bibnamefont {Rockstuhl}},
  \bibinfo {author} {\bibfnamefont {C.}~\bibnamefont {Etrich}}, \ and\ \bibinfo
  {author} {\bibfnamefont {K.}~\bibnamefont {Kern}},\ }\href {\doibase
  10.1021/nl801396r} {\bibfield  {journal} {\bibinfo  {journal} {Nano Letters}\
  }\textbf {\bibinfo {volume} {8}},\ \bibinfo {pages} {3155} (\bibinfo {year}
  {2008})}\BibitemShut {NoStop}%
\bibitem [{\citenamefont {Schnell}\ \emph {et~al.}(2010)\citenamefont
  {Schnell}, \citenamefont {Garcia-Etxarri}, \citenamefont {Alkorta},
  \citenamefont {Aizpurua},\ and\ \citenamefont {Hillenbrand}}]{Schnell2010}%
  \BibitemOpen
  \bibfield  {author} {\bibinfo {author} {\bibfnamefont {M.}~\bibnamefont
  {Schnell}}, \bibinfo {author} {\bibfnamefont {A.}~\bibnamefont
  {Garcia-Etxarri}}, \bibinfo {author} {\bibfnamefont {J.}~\bibnamefont
  {Alkorta}}, \bibinfo {author} {\bibfnamefont {J.}~\bibnamefont {Aizpurua}}, \
  and\ \bibinfo {author} {\bibfnamefont {R.}~\bibnamefont {Hillenbrand}},\
  }\href {\doibase 10.1021/nl101693a} {\bibfield  {journal} {\bibinfo
  {journal} {Nano Letters}\ }\textbf {\bibinfo {volume} {10}},\ \bibinfo
  {pages} {3524} (\bibinfo {year} {2010})}\BibitemShut {NoStop}%
\bibitem [{\citenamefont {Zhou}\ \emph
  {et~al.}(2014{\natexlab{b}})\citenamefont {Zhou}, \citenamefont {Li},\ and\
  \citenamefont {Xu}}]{Zhou2014}%
  \BibitemOpen
  \bibfield  {author} {\bibinfo {author} {\bibfnamefont {N.}~\bibnamefont
  {Zhou}}, \bibinfo {author} {\bibfnamefont {Y.}~\bibnamefont {Li}}, \ and\
  \bibinfo {author} {\bibfnamefont {X.}~\bibnamefont {Xu}},\ }\href {\doibase
  10.1364/OE.22.018715} {\bibfield  {journal} {\bibinfo  {journal} {Optics
  Express}\ }\textbf {\bibinfo {volume} {22}},\ \bibinfo {pages} {18715}
  (\bibinfo {year} {2014}{\natexlab{b}})}\BibitemShut {NoStop}%
\bibitem [{\citenamefont {Ash}\ and\ \citenamefont {Nicholls}(1972)}]{Ash1972}%
  \BibitemOpen
  \bibfield  {author} {\bibinfo {author} {\bibfnamefont {E.~A.}\ \bibnamefont
  {Ash}}\ and\ \bibinfo {author} {\bibfnamefont {G.}~\bibnamefont {Nicholls}},\
  }\href@noop {} {\bibfield  {journal} {\bibinfo  {journal} {Nature}\ }\textbf
  {\bibinfo {volume} {237}},\ \bibinfo {pages} {510} (\bibinfo {year}
  {1972})}\BibitemShut {NoStop}%
\bibitem [{\citenamefont {Lewis}\ \emph {et~al.}(1984)\citenamefont {Lewis},
  \citenamefont {Isaacson}, \citenamefont {Harootunian},\ and\ \citenamefont
  {Muray}}]{Lewis1984}%
  \BibitemOpen
  \bibfield  {author} {\bibinfo {author} {\bibfnamefont {A.}~\bibnamefont
  {Lewis}}, \bibinfo {author} {\bibfnamefont {M.}~\bibnamefont {Isaacson}},
  \bibinfo {author} {\bibfnamefont {A.}~\bibnamefont {Harootunian}}, \ and\
  \bibinfo {author} {\bibfnamefont {A.}~\bibnamefont {Muray}},\ }\href
  {\doibase 10.1017/CBO9781107415324.004} {\bibfield  {journal} {\bibinfo
  {journal} {Ultramicroscopy}\ }\textbf {\bibinfo {volume} {13}},\ \bibinfo
  {pages} {227} (\bibinfo {year} {1984})},\ \Eprint
  {http://arxiv.org/abs/arXiv:1011.1669v3} {arXiv:arXiv:1011.1669v3}
  \BibitemShut {NoStop}%
\bibitem [{\citenamefont {Pohl}\ \emph {et~al.}(1984)\citenamefont {Pohl},
  \citenamefont {Denk},\ and\ \citenamefont {Lanz}}]{Pohl1984}%
  \BibitemOpen
  \bibfield  {author} {\bibinfo {author} {\bibfnamefont {D.~W.}\ \bibnamefont
  {Pohl}}, \bibinfo {author} {\bibfnamefont {W.}~\bibnamefont {Denk}}, \ and\
  \bibinfo {author} {\bibfnamefont {M.}~\bibnamefont {Lanz}},\ }\href {\doibase
  10.1063/1.94865} {\bibfield  {journal} {\bibinfo  {journal} {Applied Physics
  Letters}\ }\textbf {\bibinfo {volume} {44}},\ \bibinfo {pages} {651}
  (\bibinfo {year} {1984})}\BibitemShut {NoStop}%
\bibitem [{\citenamefont {Betzig}\ and\ \citenamefont
  {Trautman}(1992)}]{Betzig1992}%
  \BibitemOpen
  \bibfield  {author} {\bibinfo {author} {\bibfnamefont {E.}~\bibnamefont
  {Betzig}}\ and\ \bibinfo {author} {\bibfnamefont {J.~K.}\ \bibnamefont
  {Trautman}},\ }\href {\doibase 10.1126/science.257.5067.189} {\bibfield
  {journal} {\bibinfo  {journal} {Science}\ }\textbf {\bibinfo {volume}
  {257}},\ \bibinfo {pages} {189} (\bibinfo {year} {1992})}\BibitemShut
  {NoStop}%
\bibitem [{\citenamefont {Betzig}\ and\ \citenamefont
  {Chichester}(1993)}]{Betzig1993}%
  \BibitemOpen
  \bibfield  {author} {\bibinfo {author} {\bibfnamefont {E.}~\bibnamefont
  {Betzig}}\ and\ \bibinfo {author} {\bibfnamefont {R.~J.}\ \bibnamefont
  {Chichester}},\ }\href {\doibase 10.1126/science.262.5138.1422} {\bibfield
  {journal} {\bibinfo  {journal} {Science}\ }\textbf {\bibinfo {volume}
  {262}},\ \bibinfo {pages} {1422} (\bibinfo {year} {1993})}\BibitemShut
  {NoStop}%
\bibitem [{\citenamefont {Bethe}(1944)}]{Bethe1944}%
  \BibitemOpen
  \bibfield  {author} {\bibinfo {author} {\bibfnamefont {H.~A.}\ \bibnamefont
  {Bethe}},\ }\href {\doibase 10.1103/PhysRev.66.163} {\bibfield  {journal}
  {\bibinfo  {journal} {Physical Review}\ }\textbf {\bibinfo {volume} {66}},\
  \bibinfo {pages} {163} (\bibinfo {year} {1944})}\BibitemShut {NoStop}%
\bibitem [{\citenamefont {Bouwkamp}(1950)}]{Bouwkamp1950}%
  \BibitemOpen
  \bibfield  {author} {\bibinfo {author} {\bibfnamefont {C.~J.}\ \bibnamefont
  {Bouwkamp}},\ }\href@noop {} {\bibfield  {journal} {\bibinfo  {journal}
  {Philips Research Reports}\ }\textbf {\bibinfo {volume} {5}},\ \bibinfo
  {pages} {321} (\bibinfo {year} {1950})}\BibitemShut {NoStop}%
\bibitem [{\citenamefont {Novotny}\ and\ \citenamefont
  {Hecht}(2012)}]{Novotny2012}%
  \BibitemOpen
  \bibfield  {author} {\bibinfo {author} {\bibfnamefont {L.}~\bibnamefont
  {Novotny}}\ and\ \bibinfo {author} {\bibfnamefont {B.}~\bibnamefont
  {Hecht}},\ }in\ \href@noop {} {\emph {\bibinfo {booktitle} {Principles of
  Nano-Optics}}}\ (\bibinfo  {publisher} {Cambridge University Press},\
  \bibinfo {address} {Cambridge, UK},\ \bibinfo {year} {2012})\ \bibinfo
  {edition} {2nd}\ ed.,\ Chap.~\bibinfo {chapter} {6}, pp.\ \bibinfo {pages}
  {170--183}\BibitemShut {NoStop}%
\bibitem [{\citenamefont {Wickramasinghe}\ and\ \citenamefont
  {Williams}(1990)}]{Wickramasinghe1990}%
  \BibitemOpen
  \bibfield  {author} {\bibinfo {author} {\bibfnamefont {H.~K.}\ \bibnamefont
  {Wickramasinghe}}\ and\ \bibinfo {author} {\bibfnamefont {C.~C.}\
  \bibnamefont {Williams}},\ }\href@noop {} {\enquote {\bibinfo {title}
  {{Apertureless near field optical microscope}},}\ } (\bibinfo {year}
  {1990})\BibitemShut {NoStop}%
\bibitem [{\citenamefont {Zenhausern}\ \emph {et~al.}(1995)\citenamefont
  {Zenhausern}, \citenamefont {Martin},\ and\ \citenamefont
  {Wickramasinghe}}]{Zenhausern1995}%
  \BibitemOpen
  \bibfield  {author} {\bibinfo {author} {\bibfnamefont {F.}~\bibnamefont
  {Zenhausern}}, \bibinfo {author} {\bibfnamefont {Y.}~\bibnamefont {Martin}},
  \ and\ \bibinfo {author} {\bibfnamefont {H.~K.}\ \bibnamefont
  {Wickramasinghe}},\ }\href@noop {} {\bibfield  {journal} {\bibinfo  {journal}
  {Science}\ }\textbf {\bibinfo {volume} {269}},\ \bibinfo {pages} {1083}
  (\bibinfo {year} {1995})}\BibitemShut {NoStop}%
\bibitem [{\citenamefont {Lahrech}\ \emph {et~al.}(1996)\citenamefont
  {Lahrech}, \citenamefont {Bachelot}, \citenamefont {Gleyzes},\ and\
  \citenamefont {Boccara}}]{Lahrech1996}%
  \BibitemOpen
  \bibfield  {author} {\bibinfo {author} {\bibfnamefont {A.}~\bibnamefont
  {Lahrech}}, \bibinfo {author} {\bibfnamefont {R.}~\bibnamefont {Bachelot}},
  \bibinfo {author} {\bibfnamefont {P.}~\bibnamefont {Gleyzes}}, \ and\
  \bibinfo {author} {\bibfnamefont {A.~C.}\ \bibnamefont {Boccara}},\ }\href
  {\doibase 10.1364/OL.21.001315} {\bibfield  {journal} {\bibinfo  {journal}
  {Optics Letters}\ }\textbf {\bibinfo {volume} {21}},\ \bibinfo {pages} {1315}
  (\bibinfo {year} {1996})}\BibitemShut {NoStop}%
\bibitem [{\citenamefont {Knoll}\ and\ \citenamefont
  {Keilmann}(1999)}]{Knoll1999}%
  \BibitemOpen
  \bibfield  {author} {\bibinfo {author} {\bibfnamefont {B.}~\bibnamefont
  {Knoll}}\ and\ \bibinfo {author} {\bibfnamefont {F.}~\bibnamefont
  {Keilmann}},\ }\href {\doibase 10.1038/20154} {\bibfield  {journal} {\bibinfo
   {journal} {Nature}\ }\textbf {\bibinfo {volume} {399}},\ \bibinfo {pages}
  {7} (\bibinfo {year} {1999})}\BibitemShut {NoStop}%
\bibitem [{\citenamefont {Martin}\ \emph {et~al.}(2001)\citenamefont {Martin},
  \citenamefont {Hamann},\ and\ \citenamefont {Wickramasinghe}}]{Martin2001}%
  \BibitemOpen
  \bibfield  {author} {\bibinfo {author} {\bibfnamefont {Y.~C.}\ \bibnamefont
  {Martin}}, \bibinfo {author} {\bibfnamefont {H.~F.}\ \bibnamefont {Hamann}},
  \ and\ \bibinfo {author} {\bibfnamefont {H.~K.}\ \bibnamefont
  {Wickramasinghe}},\ }\href {\doibase 10.1063/1.1354655} {\bibfield  {journal}
  {\bibinfo  {journal} {Journal of Applied Physics}\ }\textbf {\bibinfo
  {volume} {89}},\ \bibinfo {pages} {5774} (\bibinfo {year}
  {2001})}\BibitemShut {NoStop}%
\bibitem [{\citenamefont {Stockman}(2004)}]{Stockman2004}%
  \BibitemOpen
  \bibfield  {author} {\bibinfo {author} {\bibfnamefont {M.~I.}\ \bibnamefont
  {Stockman}},\ }\href {\doibase 10.1103/PhysRevLett.93.137404} {\bibfield
  {journal} {\bibinfo  {journal} {Physical Review Letters}\ }\textbf {\bibinfo
  {volume} {93}},\ \bibinfo {pages} {137404} (\bibinfo {year}
  {2004})}\BibitemShut {NoStop}%
\bibitem [{\citenamefont {Bek}\ \emph {et~al.}(2006)\citenamefont {Bek},
  \citenamefont {Vogelgesang},\ and\ \citenamefont {Kern}}]{Bek2006}%
  \BibitemOpen
  \bibfield  {author} {\bibinfo {author} {\bibfnamefont {A.}~\bibnamefont
  {Bek}}, \bibinfo {author} {\bibfnamefont {R.}~\bibnamefont {Vogelgesang}}, \
  and\ \bibinfo {author} {\bibfnamefont {K.}~\bibnamefont {Kern}},\ }\href
  {\doibase 10.1063/1.2190211} {\bibfield  {journal} {\bibinfo  {journal}
  {Review of Scientific Instruments}\ }\textbf {\bibinfo {volume} {77}},\
  \bibinfo {pages} {043703} (\bibinfo {year} {2006})}\BibitemShut {NoStop}%
\bibitem [{\citenamefont {Knoll}\ and\ \citenamefont
  {Keilmann}(2000)}]{Knoll2000}%
  \BibitemOpen
  \bibfield  {author} {\bibinfo {author} {\bibfnamefont {B.}~\bibnamefont
  {Knoll}}\ and\ \bibinfo {author} {\bibfnamefont {F.}~\bibnamefont
  {Keilmann}},\ }\href {\doibase 10.1016/S0030-4018(00)00826-9} {\bibfield
  {journal} {\bibinfo  {journal} {Optics Communications}\ }\textbf {\bibinfo
  {volume} {182}},\ \bibinfo {pages} {321} (\bibinfo {year}
  {2000})}\BibitemShut {NoStop}%
\bibitem [{\citenamefont {Bouhelier}\ \emph {et~al.}(2003)\citenamefont
  {Bouhelier}, \citenamefont {Beversluis},\ and\ \citenamefont
  {Novotny}}]{Bouhelier2003}%
  \BibitemOpen
  \bibfield  {author} {\bibinfo {author} {\bibfnamefont {A.}~\bibnamefont
  {Bouhelier}}, \bibinfo {author} {\bibfnamefont {M.~R.}\ \bibnamefont
  {Beversluis}}, \ and\ \bibinfo {author} {\bibfnamefont {L.}~\bibnamefont
  {Novotny}},\ }\href {\doibase 10.1063/1.1586482} {\bibfield  {journal}
  {\bibinfo  {journal} {Applied Physics Letters}\ }\textbf {\bibinfo {volume}
  {82}},\ \bibinfo {pages} {4596} (\bibinfo {year} {2003})}\BibitemShut
  {NoStop}%
\bibitem [{\citenamefont {Chen}\ \emph {et~al.}(2004)\citenamefont {Chen},
  \citenamefont {Chin}, \citenamefont {Ashby},\ and\ \citenamefont
  {Lieber}}]{Chen2004}%
  \BibitemOpen
  \bibfield  {author} {\bibinfo {author} {\bibfnamefont {L.}~\bibnamefont
  {Chen}}, \bibinfo {author} {\bibfnamefont {L.~C.}\ \bibnamefont {Chin}},
  \bibinfo {author} {\bibfnamefont {P.~D.}\ \bibnamefont {Ashby}}, \ and\
  \bibinfo {author} {\bibfnamefont {C.~M.}\ \bibnamefont {Lieber}},\ }\href
  {\doibase 10.1021/nl048986o} {\bibfield  {journal} {\bibinfo  {journal} {Nano
  Letters}\ }\textbf {\bibinfo {volume} {4}},\ \bibinfo {pages} {1725}
  (\bibinfo {year} {2004})}\BibitemShut {NoStop}%
\bibitem [{\citenamefont {Calleja}\ \emph {et~al.}(2002)\citenamefont
  {Calleja}, \citenamefont {Tello},\ and\ \citenamefont
  {Garc{\'{i}}a}}]{Calleja2002}%
  \BibitemOpen
  \bibfield  {author} {\bibinfo {author} {\bibfnamefont {M.}~\bibnamefont
  {Calleja}}, \bibinfo {author} {\bibfnamefont {M.}~\bibnamefont {Tello}}, \
  and\ \bibinfo {author} {\bibfnamefont {R.}~\bibnamefont {Garc{\'{i}}a}},\
  }\href {\doibase 10.1063/1.1510171} {\bibfield  {journal} {\bibinfo
  {journal} {Journal of Applied Physics}\ }\textbf {\bibinfo {volume} {92}},\
  \bibinfo {pages} {5539} (\bibinfo {year} {2002})}\BibitemShut {NoStop}%
\end{thebibliography}%


\renewcommand{\thefigure}{S\arabic{figure}}

\setcounter{figure}{0}

\section{\label{sec:SIsSNOM}Supplemental Information}
This supplemental information contains additional modeled near-field mappings as well as other graphs to more fully show how the modeled results change with different values used for the input parameters (tip radius, distance of closest approach – DCA) and support paper conclusions. 

\subsection{\label{sec:sSNOMparam}Additional Mappings for the Parameters Chosen in the Main Text (15 nm radius, 8 nm DCA)}

\newpage
\begin{figure}[H]
\begin{centering}
	\includegraphics[width=0.9\textwidth]{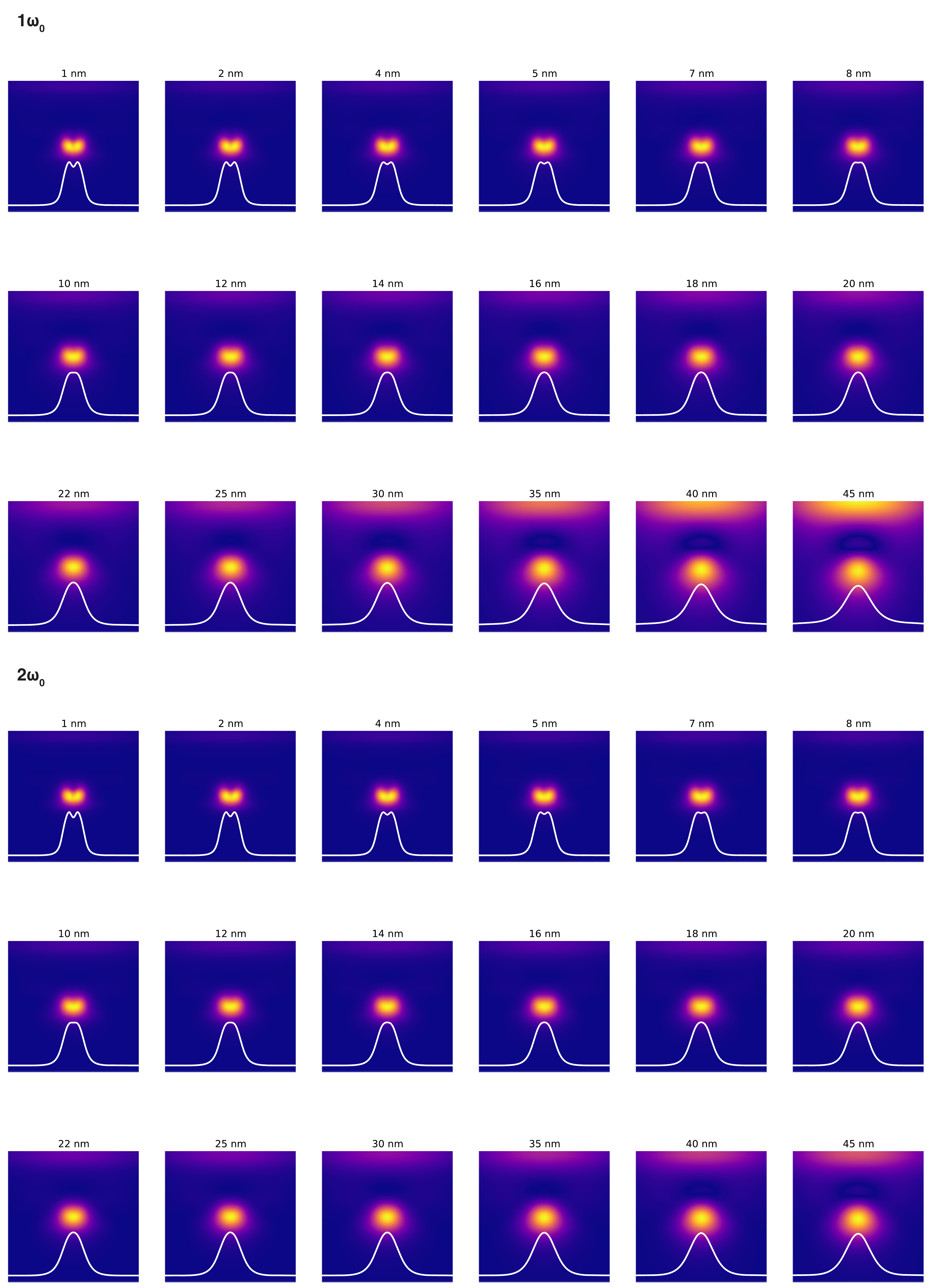}
	\caption[Complete harmonics for Figure \ref{figure4}]{\textbf{Complete harmonics for Figure \ref{figure4}.} (Next 2 pages also.) The set of six harmonic mappings complementary to Figure \ref{figure4} where only $1 \omega _0$ and $6 \omega _0$ were given.}
	\label{SIHarm815}
\end{centering}
\end{figure}

\begin{figure}[H]
\begin{center}
	\includegraphics[width=0.9\textwidth]{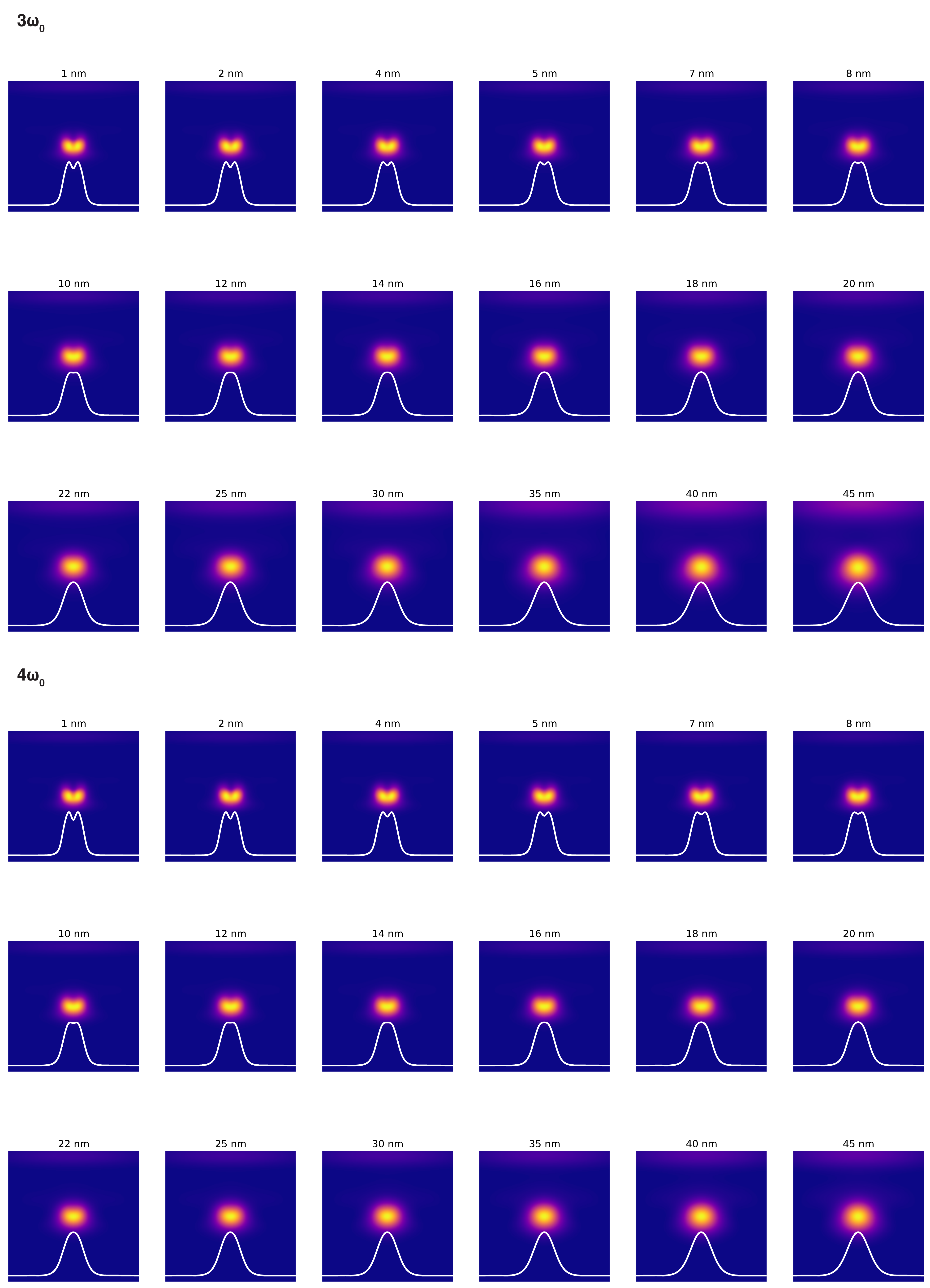}
\end{center}
\end{figure}

\begin{figure}[H]
\begin{center}
	\includegraphics[width=0.9\textwidth]{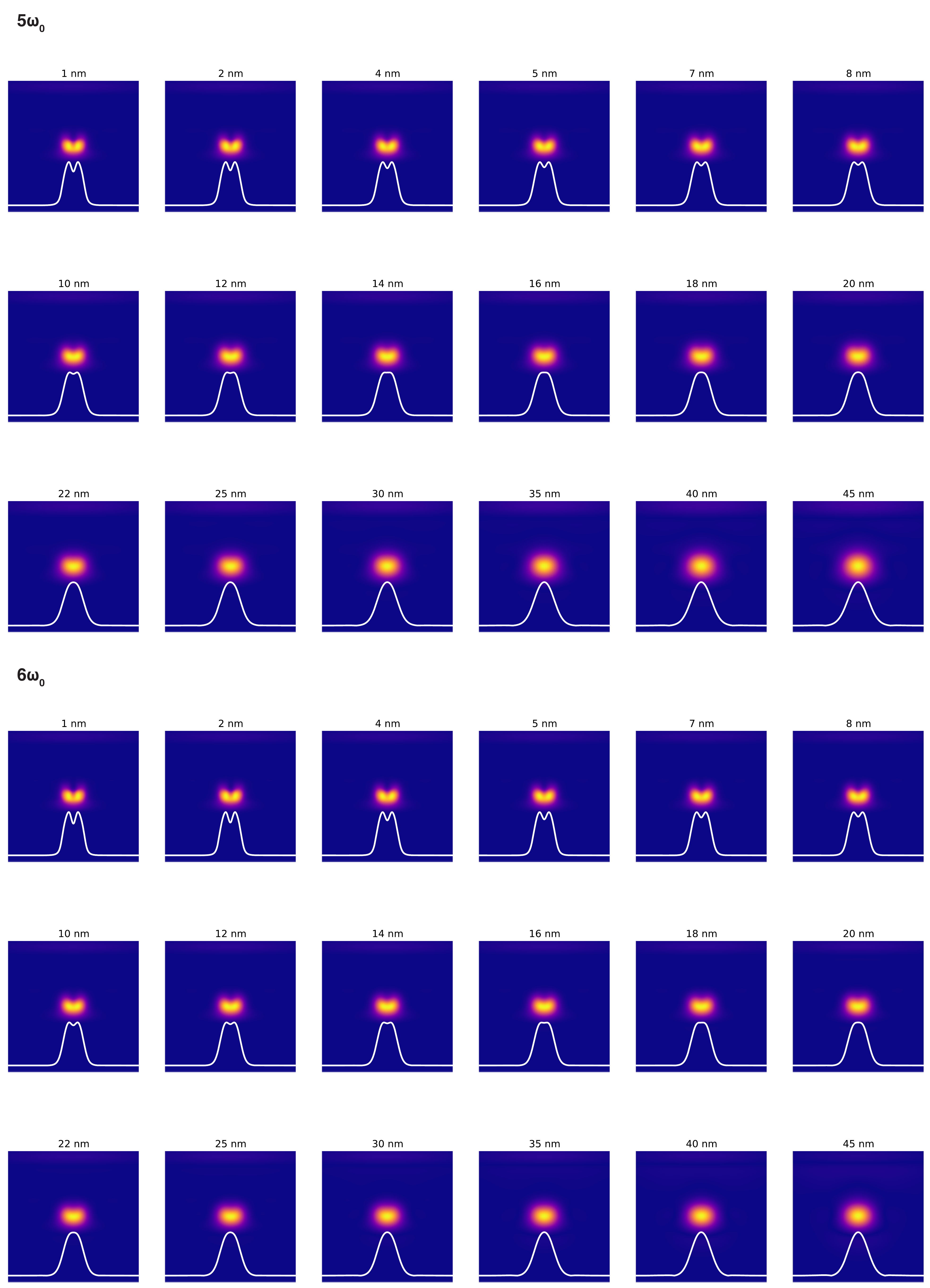}
\end{center}
\end{figure}

\subsection{\label{sec:sSNOMparamTrend}Figures for Observing Trends in the Maximums of Each Mapping as a Function of DCA Height, Tip Radius, and Harmonic}

\begin{figure}[H]
\begin{centering}
	\includegraphics[width=1\textwidth]{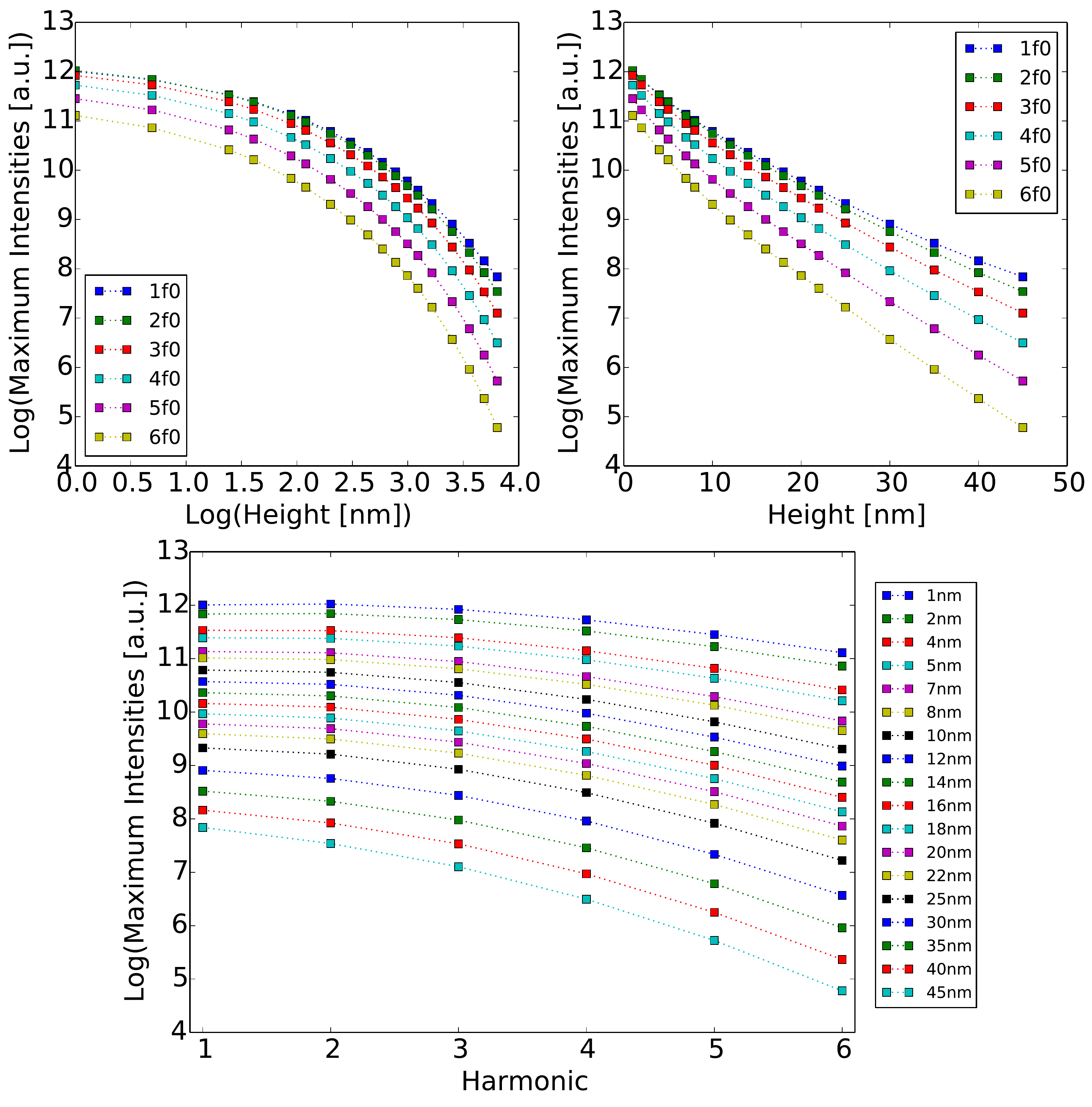}
	\caption[Maximum intensity decay with DCA and harmonic]{\textbf{Maximum intensity decay with DCA and harmonic.} The decay relationship of the maximum intensity of each mapping as a function of both DCA height and harmonic. Faster decay is observed for higher DCAs as well as for higher harmonics.}
	\label{SIdecay}
\end{centering}
\end{figure}

\begin{figure}[H]
\begin{centering}
	\includegraphics[width=0.5\textwidth]{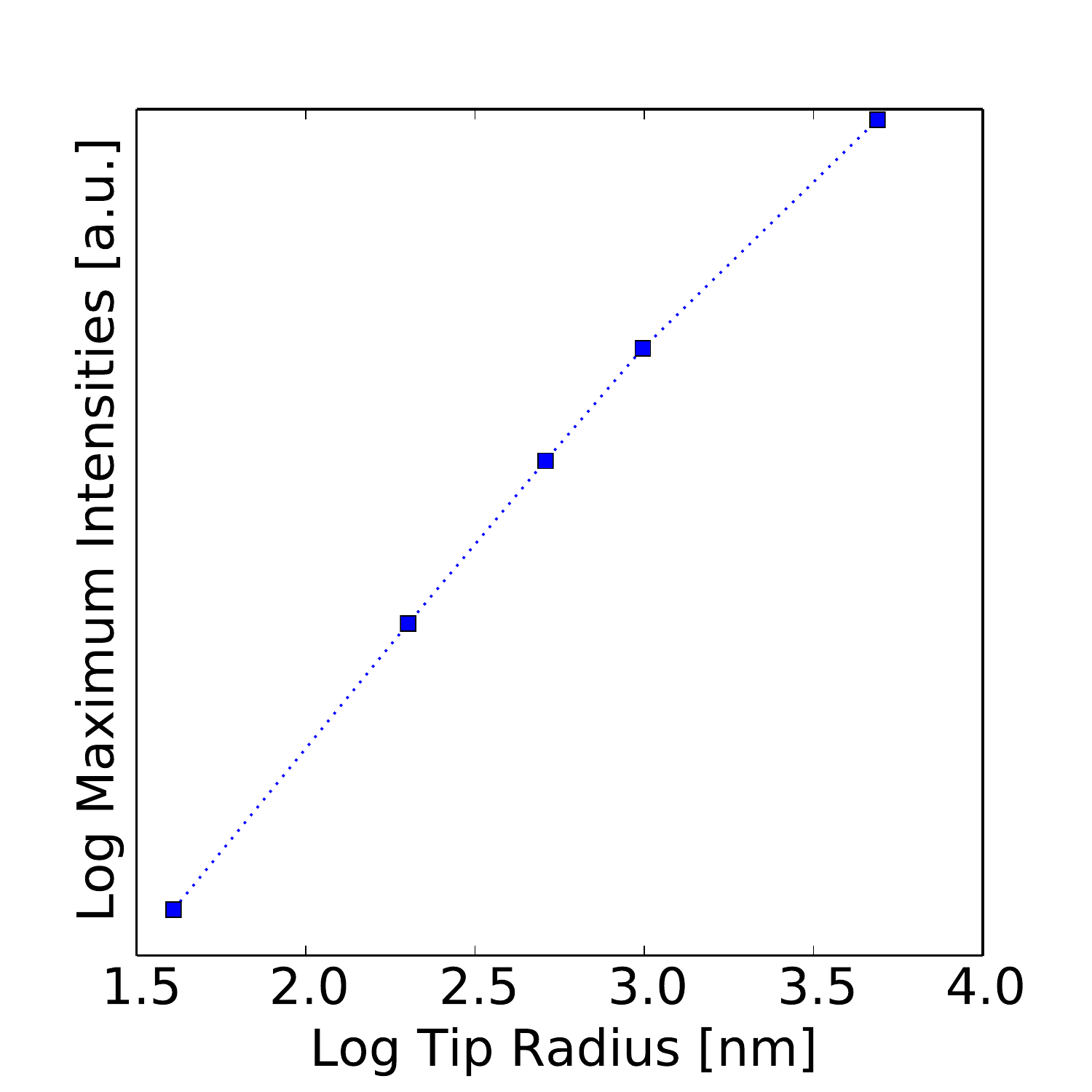}
	\caption[Maximum intensities vs. tip radius]{\textbf{Maximum intensities vs. tip radius.} The maximum of each mapping as a function of modeled tip radius plotted for only the $1 \omega _0$ but representative of all harmonics.}
	\label{SIMItiprad}
\end{centering}
\end{figure}

\subsection{\label{sec:sSNOMrad}Near-field Mappings for Other Modeled Tip Radii}

\begin{figure}[H]
\begin{centering}
	\caption[Harmonics for a 20 nm tip radius]{\textbf{Harmonics for a 20 nm tip radius.} (Next page.) The modeled $1 \omega _0$ and $6 \omega _0$ for a 20 nm tip radius. A tip radius of this size was determined to be too large to fit the experimentally measured data. The near-field mappings generated by the model for the $6 \omega _0$ do not show a bifurcation in the shape of the near-field spot as is seen in the experimental data. This is because the top-hat shaped tip with a 20 nm radius is now comparable in size to the plasmonic antenna and its near-field spot. The tip simply scatters the field from the two corner spots out across the whole antenna as it scans over, which smears out the observed response.}
	\label{SIHarm20}
\end{centering}
\end{figure}

\begin{figure}[H]
\begin{center}
	\includegraphics[width=0.85\textwidth]{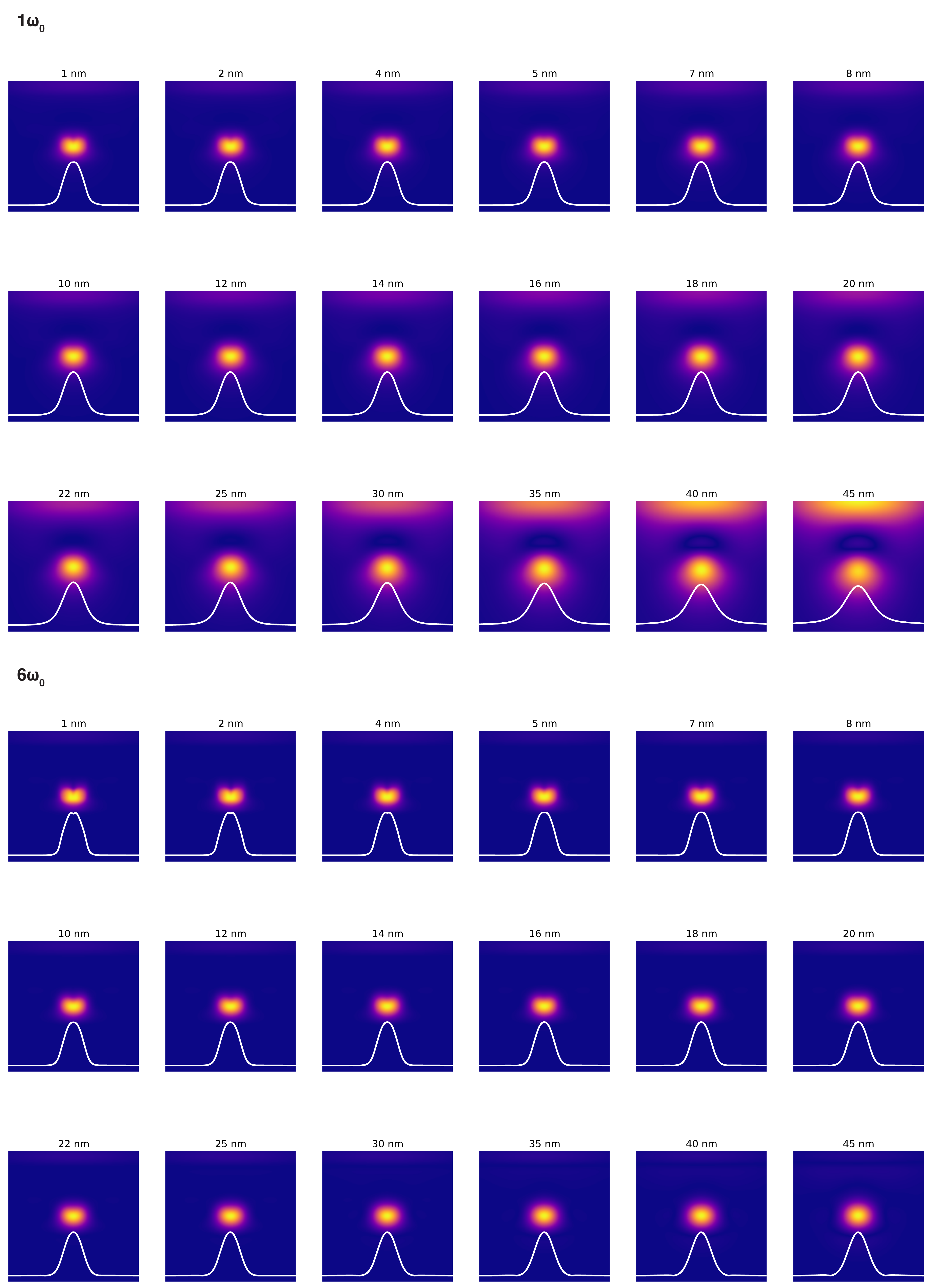}
\end{center}
\end{figure}

\begin{figure}[H]
\begin{centering}
	\includegraphics[width=0.85\textwidth]{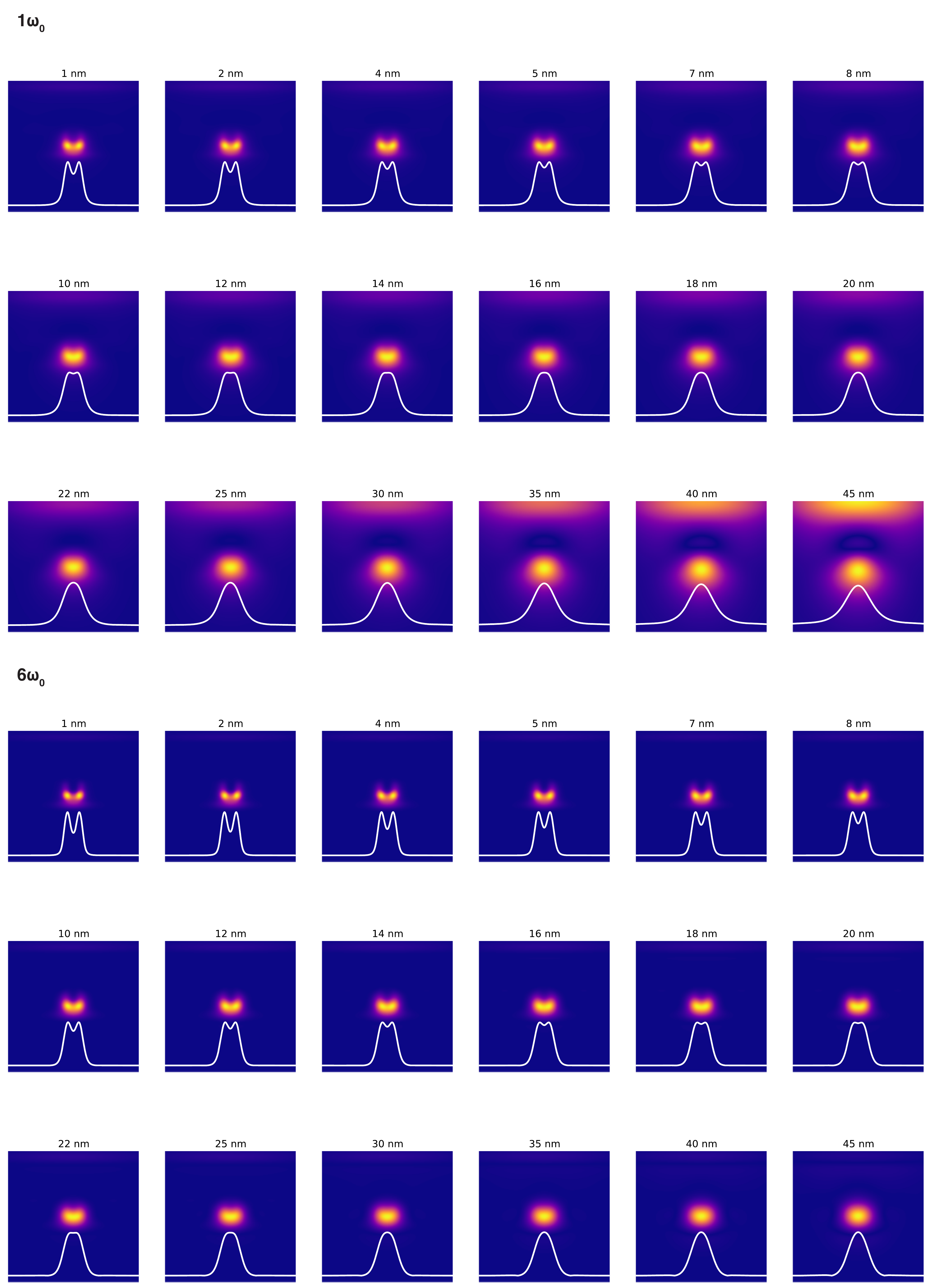}
	\caption[Harmonics for a 10 nm tip radius]{\textbf{Harmonics for a 10 nm tip radius.} The modeled $1 \omega _0$ and $6 \omega _0$ for a 10 nm tip radius. This modeled tip radius together with a DCA of $\sim$16 nm shows reasonable agreement with the experimental data; however, based on AFM mechanics, we believed that this DCA was too large to be a reasonable estimate for our experimental conditions. Additional mappings for all harmonics modeled from this condition are in Figure \ref{SIHarm1610}.}
	\label{SIHarm10}
\end{centering}
\end{figure}

\begin{figure}[H]
\begin{centering}
	\includegraphics[width=0.75\textwidth]{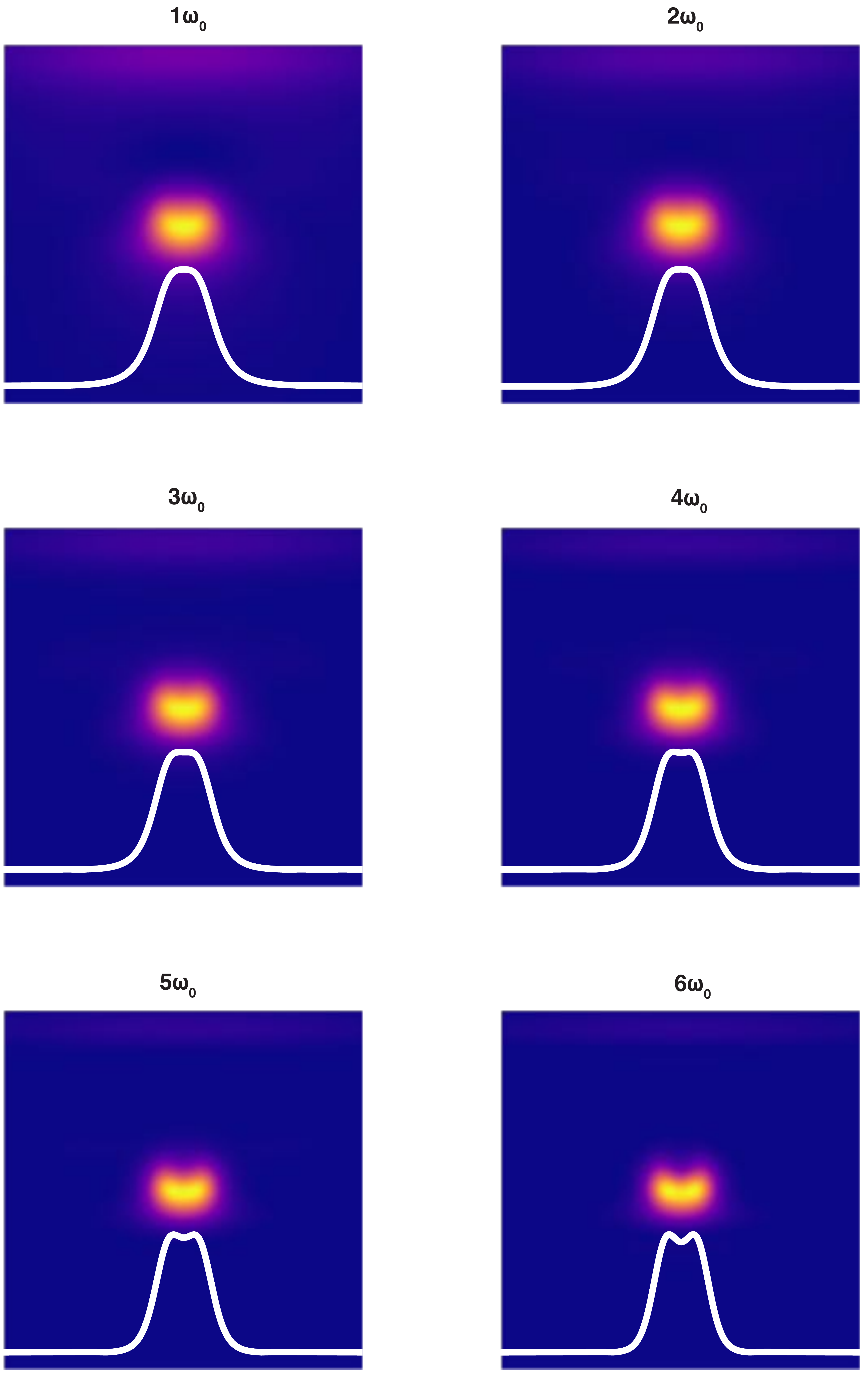}
	\caption[Harmonics for a 10 nm tip radius and 16 nm DCA]{\textbf{Harmonics for a 10 nm tip radius and 16 nm DCA.} All six modeled harmonic mappings for the 10 nm tip radius and 16 nm DCA condition.}
	\label{SIHarm1610}
\end{centering}
\end{figure}

\begin{figure}[H]
\begin{centering}
	\includegraphics[width=0.85\textwidth]{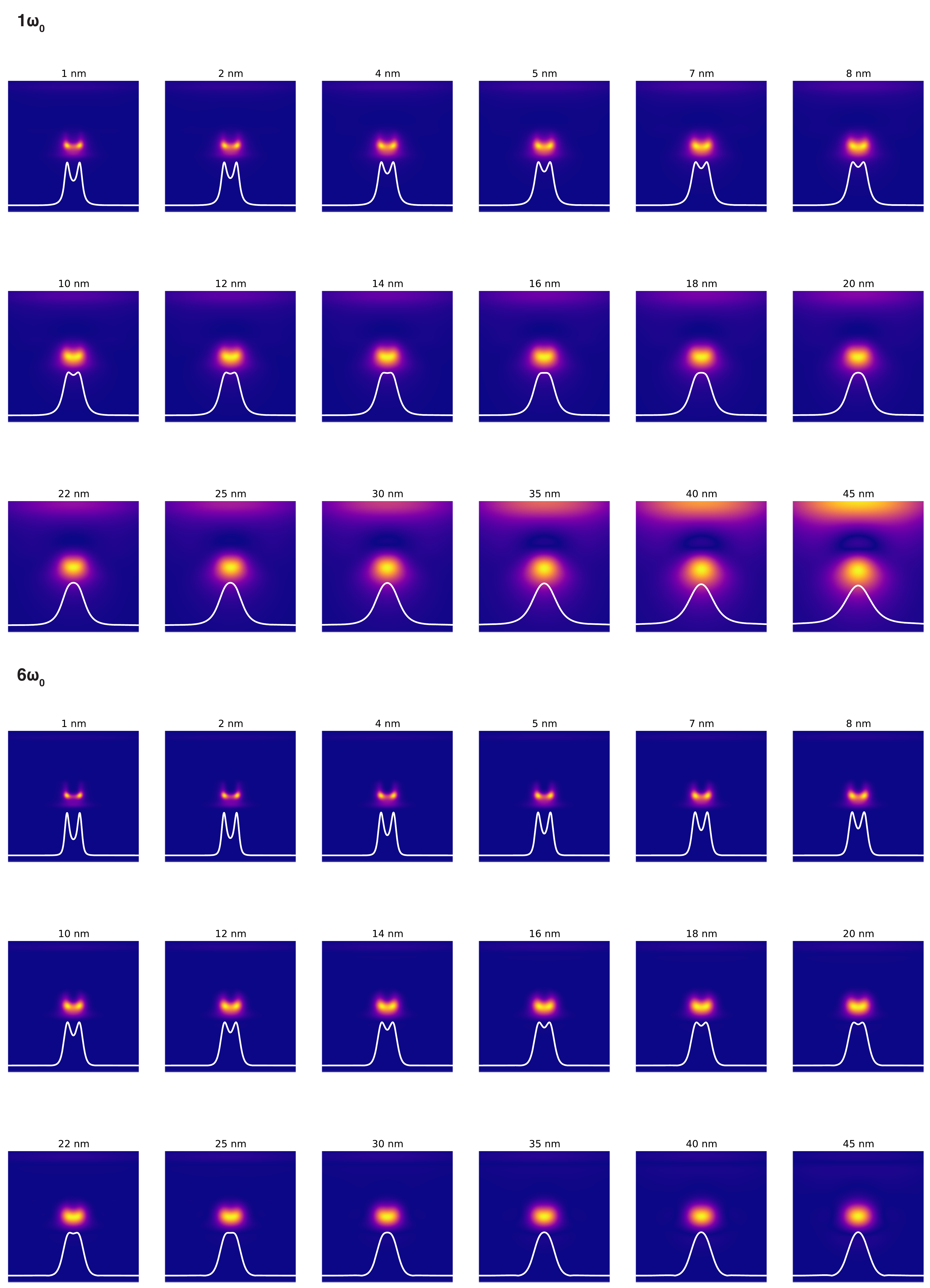}
	\caption[Harmonics for a 5 nm tip radius]{\textbf{Harmonics for a 5 nm tip radius.} The modeled $1 \omega _0$ and $6 \omega _0$ for a 5 nm tip radius.}
	\label{SIHarm5}
\end{centering}
\end{figure}

\subsection{\label{sec:sSNOMdca}Near-field Mappings for Other Modeled Distances of Closest Approach}

\begin{figure}[H]
\begin{centering}
	\includegraphics[width=0.85\textwidth]{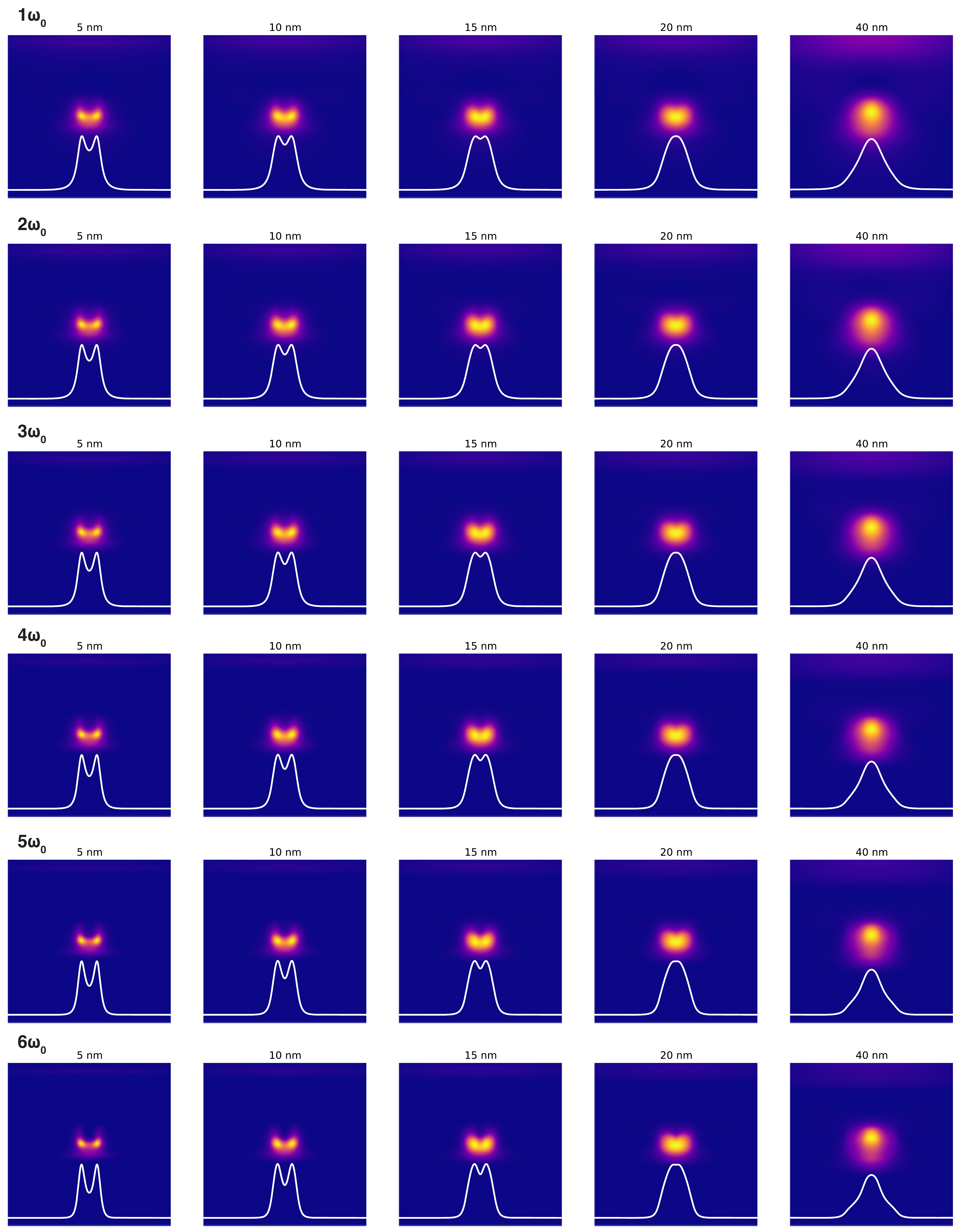}
	\caption[Harmonics for a 4 nm DCA]{\textbf{Harmonics for a 4 nm DCA.} A modeled 4 nm tip DCA shows the spot bifurcation in the $1 \omega _0$ with the 15 nm tip radius, which is not in agreement with the experimental data and the AFM mechanical expectations, so these parameters were not selected for the final model.}
	\label{SIHarm4nmDCA}
\end{centering}
\end{figure}

\begin{figure}[H]
\begin{centering}
	\includegraphics[width=0.85\textwidth]{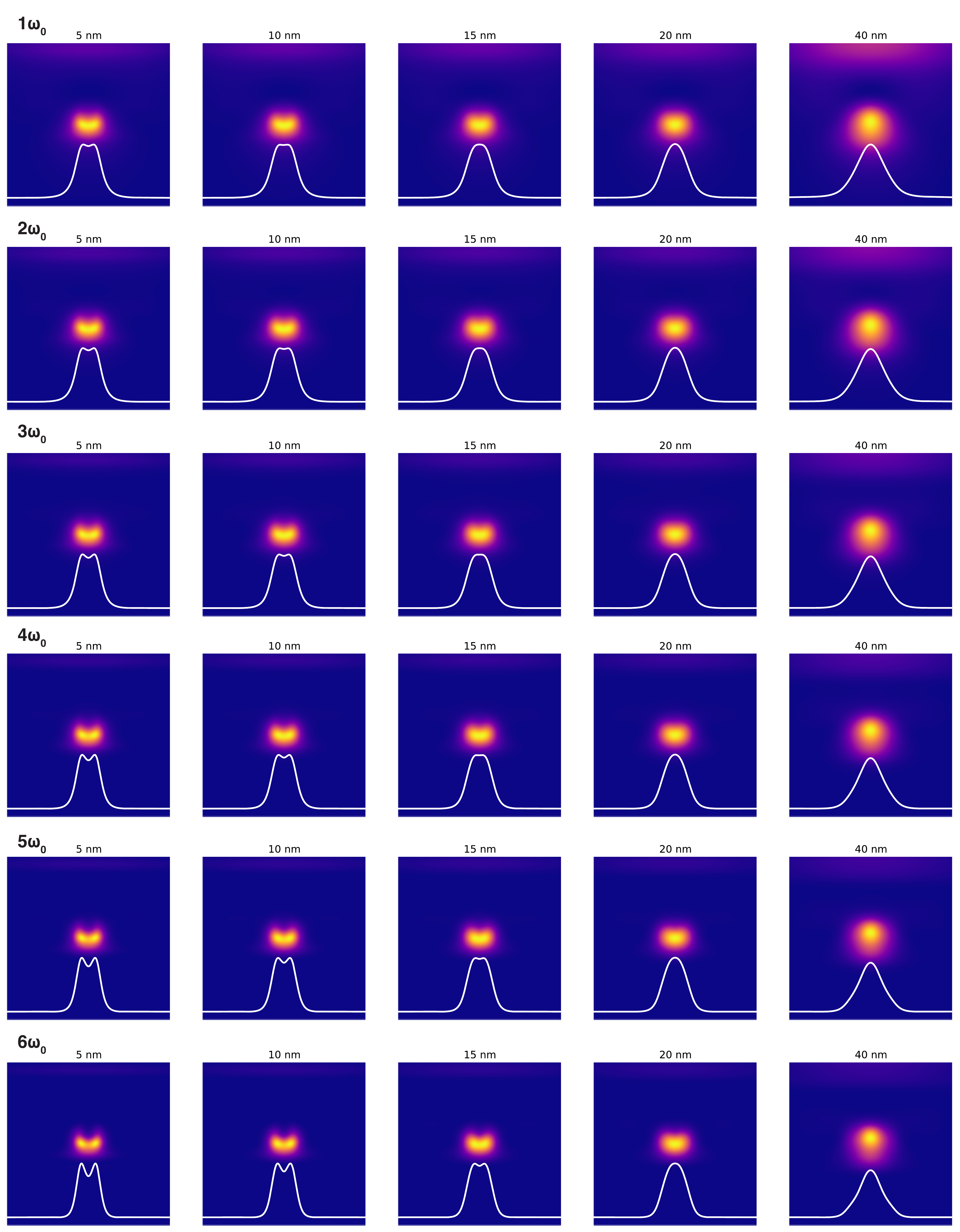}
	\caption[Harmonics for a 12 nm DCA]{\textbf{Harmonics for a 12 nm DCA.} A modeled 12 nm tip DCA does not show enough spot bifurcation in the $6 \omega _0$ with the 15 nm tip radius, and it shows too much in the $1 \omega _0$ with the 10 nm tip radius.}
	\label{SIHarm12nmDCA}
\end{centering}
\end{figure}

\subsection{\label{sec:sSNOMefield}Simulated Electric Field Intensity}

\begin{figure}[H]
\begin{centering}
	\includegraphics[width=1\textwidth]{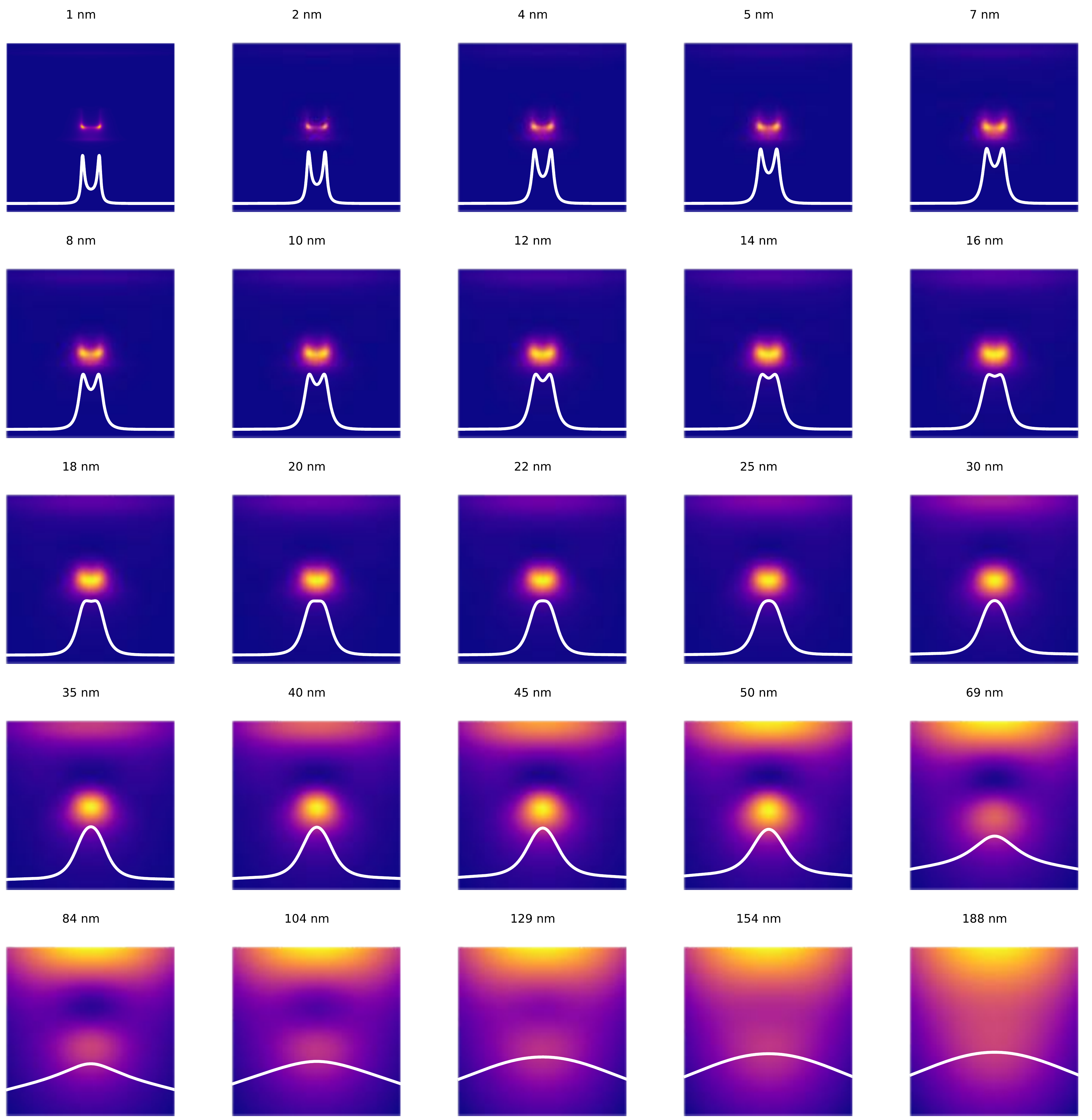}
	\caption[Simulated electric field intensity of a HAMR head]{\textbf{Simulated electric field intensity.} Simulated electric field intensity in discrete planes above the air-bearing surface (ABS) help us observe which regions play dominant roles in each harmonic mapped by the sSNOM model as well as the experimental sSNOM measurements. Maps are labeled with their height above the ABS.}
	\label{SIE2}
\end{centering}
\end{figure}

\begin{figure}[H]
\begin{centering}
	\includegraphics[width=1\textwidth]{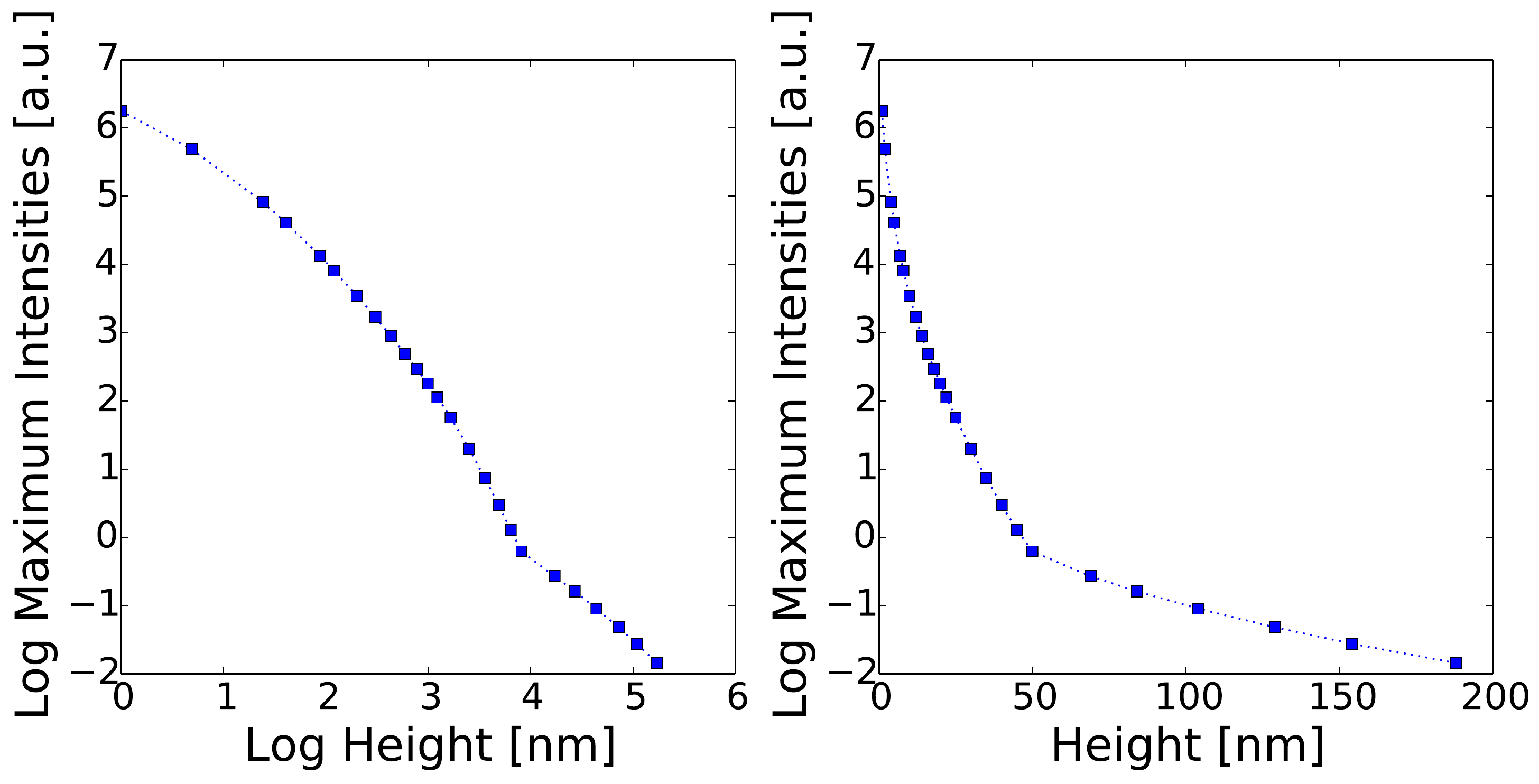}
	\caption[Maximum electric field intensities]{\textbf{Maximum electric field intensities of the maps in Figure \ref{SIE2} over the entire 400 nm $\times$ 400 nm region.} At a height of $\sim$50 nm above the ABS, the maximum electric field intensity region shifts from over the ``notch'' region of the plasmonic antenna to over the side of the ``E'' shaped antenna as is evident in Figure S10. This effect is seen in the fundamental harmonic ($1 \omega 0$) of the experimental data, and to some degree in the modeled data as well, although it is not as evident in the specific condition (15 nm tip radius, 8 nm DCA) presented in the main text.}
	\label{SIE2max}
\end{centering}
\end{figure} 

\begin{figure}[H]
\begin{centering}
	\includegraphics[width=1\textwidth]{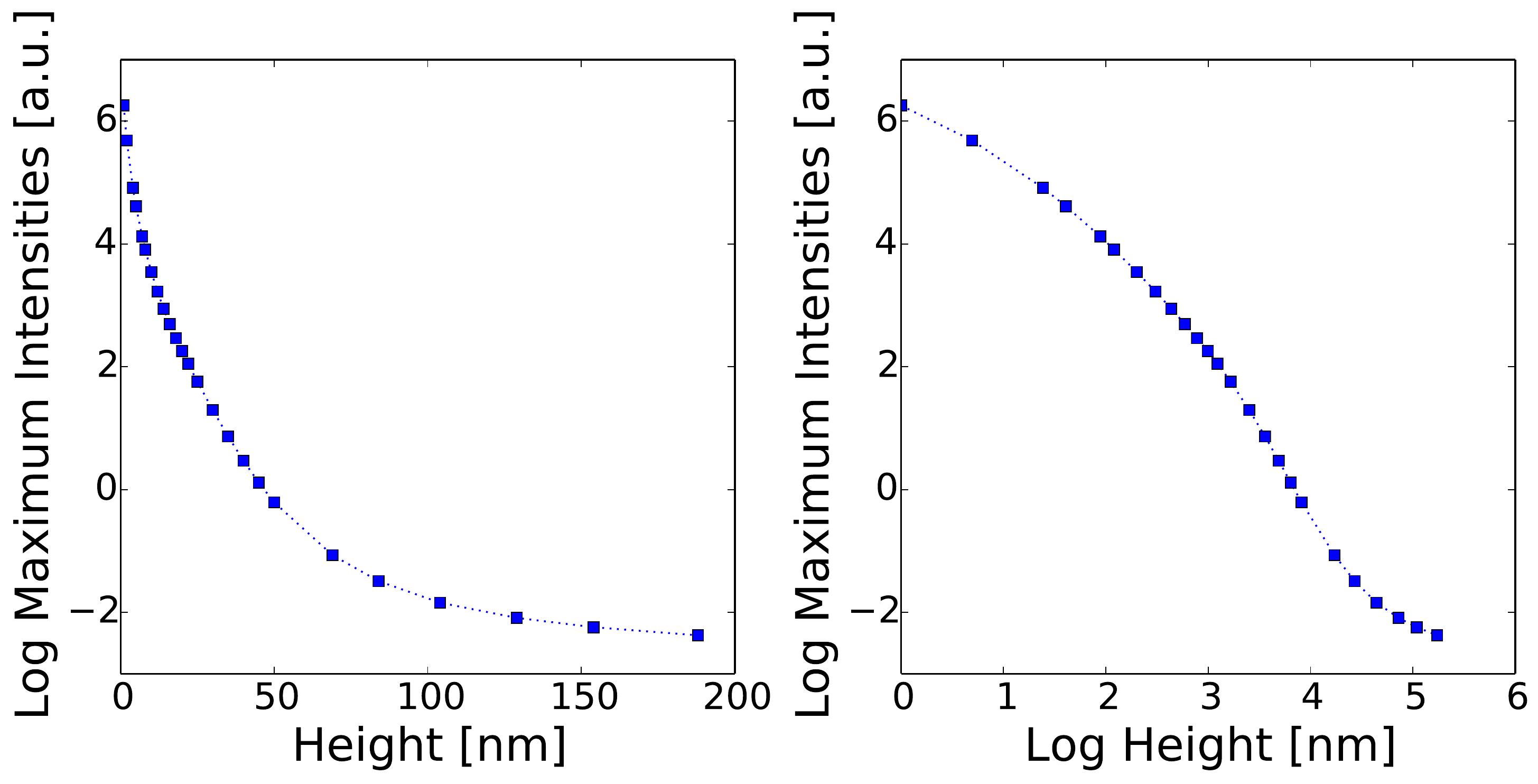}
	\caption[Maximum electric field intensities]{\textbf{Maximum electric field intensities of the maps in Figure \ref{SIE2} over the center region (over the notch that generates the main near-field spot).} The same sharp transition at $\sim$50 nm above the ABS is not as evident as in Figure \ref{SIE2max} (over the entire map shown); however, the decay trend still alters.}
	\label{SIE2maxCenter}
\end{centering}
\end{figure} 

\end{document}